\documentclass[useAMS,usenatbib]{mn2e}

\usepackage{natbib}
\usepackage{graphicx}
\usepackage{dcolumn}
\usepackage{tikz}
\usepackage{pgf}
\usepackage{bm}
\usepackage{amsmath}
\usepackage{amssymb}
\usepackage{hyperref}

\graphicspath{{figs/pdf/}{figs/jpg/}}

\newcommand{\vect}[1]{\bm{#1}}

\newcommand{\Aq}[1]{\texttt{Aq-#1}}

\title[The shape of the Aquarius dark matter haloes]{The Shape of Dark
  Matter Haloes in the Aquarius Simulations: Evolution and Memory}

\author[Vera-Ciro et al.]{
\parbox[t]{\textwidth}{
Carlos A. Vera-Ciro\textsuperscript{1}\thanks{E-mail: cavera@astro.rug.nl}, 
Laura V. Sales\textsuperscript{1,4},
Amina Helmi\textsuperscript{1},
Carlos S. Frenk\textsuperscript{2},\\
Julio F. Navarro\textsuperscript{3},
Volker Springel\textsuperscript{5,6},
Mark Vogelsberger\textsuperscript{7}
and Simon D. M. White\textsuperscript{4}
}\\ 
\\
\\
\textsuperscript{1} Kapteyn Astronomical Institute, Univ. of
Groningen, P.O. Box 800, 9700 AV Groningen, The Netherlands \\
\textsuperscript{2} Institute for Computational Cosmology, Dep. of Physics,
Univ. of Durham, South Road, Durham DH1 3LE, UK \\
\textsuperscript{3} Dep. of Physics \& Astron., University of Victoria, Victoria,
BC, V8P 5C2, Canada \\
\textsuperscript{4} Max-Plank-Institut f\"ur Astrophysik,
Karl-Schwarzschild-Stra\ss e, 1, 85740 Garching bei M\"unchen, Germany
\\
\textsuperscript{5} Heidelberg Institute for Theoretical Studies,
Schloss- Wolfsbrunnenweg 35, 69118 Heidelberg, Germany\\
\textsuperscript{6} Zentrum f\"ur Astronomie der Universit\"at
Heidelberg, Astronomisches Recheninstitut, M\"{o}nchhofstr. 12-14,
69120 Heidelberg, Germany\\
\textsuperscript{7} Harvard-Smithsonian Center for Astrophysics, 60
Garden St., Cambridge, MA 02138, USA
}

\begin{document}

\date{8 April 2011}
\pagerange{\pageref{firstpage}--\pageref{lastpage}} \pubyear{2011}

\maketitle

\label{firstpage}

\begin{abstract}

  We use the high resolution cosmological $N$-body simulations from
  the Aquarius project to investigate in detail the mechanisms that
  determine the shape of Milky Way-type dark matter haloes. We find
  that, when measured at the instantaneous virial radius, the shape of
  individual haloes changes with time, evolving from a typically
  prolate configuration at early stages to a more triaxial/oblate
  geometry at the present day. This evolution in halo shape correlates
  well with the distribution of the infalling material: prolate
  configurations arise when haloes are fed through narrow filaments,
  which characterizes the early epochs of halo assembly, whereas
  triaxial/oblate configurations result as the accretion turns more
  isotropic at later times.  Interestingly, at redshift $z=0$, clear
  imprints of the past history of each halo are recorded in their
  shapes at different radii, which also exhibit a variation from
  prolate in the inner regions to triaxial/oblate in the
  outskirts. Provided that the Aquarius haloes are fair representatives
  of Milky Way-like $10^{12}$M$_\odot$ objects, we conclude that the
  shape of such dark matter haloes is a complex, time-dependent
  property, with each radial shell retaining memory of the conditions
  at the time of collapse.

\end{abstract}


\section{Introduction}

In our current understanding of the Universe, dark matter haloes
constitute an integral part of galaxies. Their properties, especially
their density profile and shape, have received significant attention
in recent years as they have been argued to be sensitive to the
fundamental properties of the dark matter particles. Numerical
simulations have been extensively used to study the characteristics of
the dark matter haloes, exploring for example the effects of the
environment, mass assembly history and the nature of dark matter
itself \citep[e.g., ][]{Bullock2002, Bett2007, Maccio2007, Hahn2007,
Spergel2000, Yoshida2000, Avila2001,Strigari2007, Wang2009}.


The first fully analytical models of the formation of dark matter
haloes such as the top-hat spherical collapse model
\citep{Gunn_Gott1972,Fillmore1984}, considered highly symmetric
configurations. However, the pioneering work of \citet{Frenk1988,
Dubinski1991, Warren1992, Cole1996, Thomas1998} demonstrated important
deviations from spherical symmetry by measuring the shape of dark
matter haloes in numerical $N$-body simulations evolved in a fully
cosmological context. These authors consistently found that after
virialization, dark matter haloes are triaxial with more prolate
shapes towards the centre and more oblate shapes in the
outskirts. Recent high resolution $N$-body simulations have yielded
similar conclusions \citep{Jing2002, Bailin2005, Kasun2005,
Hopkins2005, Bett2007, Hayashi2007, Kuhlen2007, Stadel2009,
Diemand2009}.

Further studies based on numerical simulations have also revealed that
the environment and mass assembly history of a halo may play a crucial
role in determining its shape. Pioneering work by \citet{Tormen1997}
and \citet{Colberg1999} suggested that the anisotropic infall of
matter onto cluster-sized haloes was largely responsible for their
shape, orientations and dynamics at different times. Because infall is
governed by the surrounding large scale structure, we expect
significant correlations between the halo shapes and their
environment, although evidence both against and in support of such
trends have been reported so far in the literature \citep{Lemson1999,
Avila2005, Faltenbacher2005, Altay2006, Basilakos2006, Gottlober2006,
Patiri2006, Aragon2007b, Hahn2007b, Maccio2007, Ragone2010}.

The observational determination of the shapes of dark matter haloes is
challenging. Preferably dynamical tracers at large radii are to be
used, but these tracers are by definition rare. In external galaxies,
constraints on these shapes have been put using the intrinsic shape of
galactic discs \citep{Fasano1993}, the kinematics and morphology of
the H\textsc{i} layer \citep{Olling1996, Becquaert1997, Swaters1997},
the morphology, temperature profile of X-ray isophotes
\citep{Buote1998, Buote2002}, gravitational lensing
\citep{Hoekstra2004} and the spatial distribution of galaxies within
groups \citep{Robotham2008} \citep[for earlier reviews on the subject
see][]{Rix1996, Sackett1999}. The general consensus of all these
studies is that haloes tend to be roughly oblate, with the smallest
axis pointing perpendicular to the symmetry plane defined by the
stellar component. Most of these constrains, however, pertain to the
inner regions (a few optical radii) of galaxy-scale haloes.

In the case of the Milky-Way, the shape constraints often rely on the
kinematics of individual stars, and include e.g.\ the use of the tilt
of the velocity ellipsoid for nearby stars \citep{Siebert2008}, the
proper motions of hypervelocity stars \citep{Gnedin2005} or the
dynamics of stellar streams \citep{Koposov2010}. Interestingly, the
use of the latter have provided contradictory results, notably in the
case of the Sagittarius Stream.  For example, the positional
information was used to argue that the Milky Way halo is nearly
spherical \citep{Ibata2001,Martinez2004,Johnston2005} whereas the
kinematics of stars in the leading stream could only be fit in a
prolate halo elongated perpendicular to the disc \citep{Helmi2004,
Law2005}. More recently, \citet{Law2009, Law2010} have explored
triaxial potentials with constant axis ratios, and found models that
were able to fit simultaneously these constraints, albeit not
completely satisfactorily.

From the theoretical perspective, there is a gap in our knowledge
about the way in which dark matter haloes acquire their shape, and in
particular, on the impact of the dynamics of the surrounding large
scale structure in the non-linear regime \citep{Lee2005,
Betancort2009, Rossi2010, Salvador2011}. Indeed, most of the
theoretical works cited above restrict their analysis to the present
day correlations with the environment, and do not consider when the
shapes have been established and how they relate to the past history
of an object.  In this paper we use state-of-the-art high resolution
$N$-body simulations that track the formation of five $\sim
10^{12}$M$_\odot$ Milky Way-like haloes in a fully cosmological
context, in order to gain further understanding of this problem.

This paper is organized as follows. In
Section~\ref{sec:numerical-simulations} we describe the simulations
used in this work which are part of the Virgo Consortium's {\it
Aquarius} project, and test in Section~\ref{sec:methods} the
convergence of halo shapes for different resolutions.  The shapes at
the present day, evolution and their relation with the formation
history of the dark matter haloes are analyzed in
Section~\ref{sec:shape} and \ref{sec:lss}.  Finally, we discuss and
summarize our main conclusions in Section \ref{sec:conclusions}. For
completeness, we compare in Appendix \ref{sec:methods-appendix} the
results obtained by using several different schemes to determine halo
shapes. Also a novel method for measuring the size of the filaments
based on dynamical arguments is described in Appendix
\ref{sec:filament_size}.

\section{Numerical Simulations}
\label{sec:numerical-simulations}

We use the Aquarius Simulations, a suite of high resolution $N$-body
cosmological simulations of six Milky Way sized dark matter haloes
\citep{Springel2008}. These haloes were selected from a larger
$\Lambda$CDM cosmological box of $100 h^{-1}$ Mpc side with parameters
$\Omega_m=0.25$, $\Omega_{\Lambda} = 0.75$, $\sigma_8=0.9$, $n_s=1$
and $H_0=100 h$ km s\textsuperscript{-1}Mpc\textsuperscript{-1} $ =
73$ km s\textsuperscript{-1}Mpc\textsuperscript{-1}. The Aquarius
haloes, labeled \Aq{A} to \Aq{F}, have a final mass $\sim
10^{12}$M$_\odot$ and were chosen to be relatively isolated at
redshift $z=0$. The selection procedure was otherwise random, which
allows us to study the impact of different assembly histories on the
shape of Milky Way sized dark matter haloes.\\

Each halo has been re-simulated at various resolution levels that
accurately replicate the power-spectrum and phases for the resolved
structures in all runs. Following the notation introduced in previous
papers, we refer to each level of resolution as -5 to -1, for the
lowest to highest resolution. The mass per particle varies from $m_p =
2.94\times 10^6 h^{-1}$M$_\odot$ in the level 5 to $m_p = 1.25\times
10^3 h^{-1}$M$_\odot$ for the highest resolution run, which resolves
a given halo with approximately half-million up to 1.5-billion
particles within the virial radius for level 5 and 1,
respectively\footnote{Throughout this paper we refer to the virial
radius, $r_{\rm vir}$, as the spherical radius that contains a mean
density equal to 200 times the critical density of the Universe at a
given time. Other virial quantities, such as mass and velocities,
$m_{\rm vir}$ and $v_{\rm vir}$ respectively, correspond to those
measured within $r_{\rm vir}$.}. The results discussed in this paper
pertain mostly to the 4th level of resolution ($m_p\sim 2\times
10^5h^{-1}$M$_\odot$, $n_{\rm vir}\sim 10^{6}$ particles within
$r_{\rm vir}$, gravitational softening $\epsilon = 250h^{-1}$ pc) of
haloes \Aq{A} to \Aq{E} and aim to target the possible structure of the
Milky Way dark matter halo. The Aquarius halo \Aq{F} is not considered
in this work because it experiences a major merger less than $\sim 5$
Gyr ago; this is unlikely to be consistent with our current view of
the assembly of the Milky Way halo, expected to have had no major
mergers since $z\sim1$ \citep{Toth1992}.

In this context, this paper focuses on dark matter haloes with a
relatively quiescent merger history (mass ratios lower than 1:5) after
$z \sim 2$ \citep[for a detailed comparison of the Aquarius haloes'
mass accretion history with respect to the general expectations for
haloes of similar virial mass see][]{Boylan2010}. We summarize the most
relevant numerical details of our haloes at level 4 in Table
\ref{aquarius-details-table}, and we refer the interested reader to
\citet{Springel2008} for further information.

\begin{table}
  \begin{center}
    \begin{tabular}{cccccc} \hline\hline
      Halo & $m_p$  & $r_{\rm vir}$ &  $m_{\rm vir}$ & $n_{\rm vir}/10^6$ & $r_{\rm conv}$\\
      \hline
      \Aq{A-4} & $2.87$ & $179.36$ & $1.34$ & $4.68$ & $2.27$ \\
      \Aq{B-4} & $1.64$ & $137.86$ & $0.61$ & $3.72$ & $2.14$ \\
      \Aq{C-4} & $2.35$ & $177.89$ & $1.31$ & $5.58$ & $2.07$ \\
      \Aq{D-4} & $1.95$ & $177.83$ & $1.31$ & $6.69$ & $2.14$ \\
      \Aq{E-4} & $1.90$ & $155.95$ & $0.88$ & $4.64$ & $2.05$ \\ \hline
    \end{tabular}
  \end{center}
  \caption{\label{aquarius-details-table} Numerical details of the
    Aquarius haloes for resolution level-4. Each column lists: (1) the
    name of the halo, (2) mass per particle $m_p$ in
    $10^5$M$_{\odot}\ h^{-1}$ , (3) the virial radius $r_{\rm vir}$ in
    $\mathrm{kpc}\ h^{-1}$ and (4) virial mass $m_{\rm vir}$ in
    $10^{12}$M$_{\odot}\ h^{-1}$ at $z=0$, (5) $n_{\rm vir}$ the
    number of dark matter particles within $r_{\rm vir}$, (6) $r_{\rm
      conv}$ the convergence radius in ${\rm kpc}\ h^{-1}$. The
    gravitational softening length is $\epsilon = 250h^{-1}$ pc for
    all haloes at this resolution level.}
\end{table}

\section{Halo shape determination and convergence}
\label{sec:methods}

\begin{figure}
  \centering \includegraphics[width =
  0.45\textwidth]{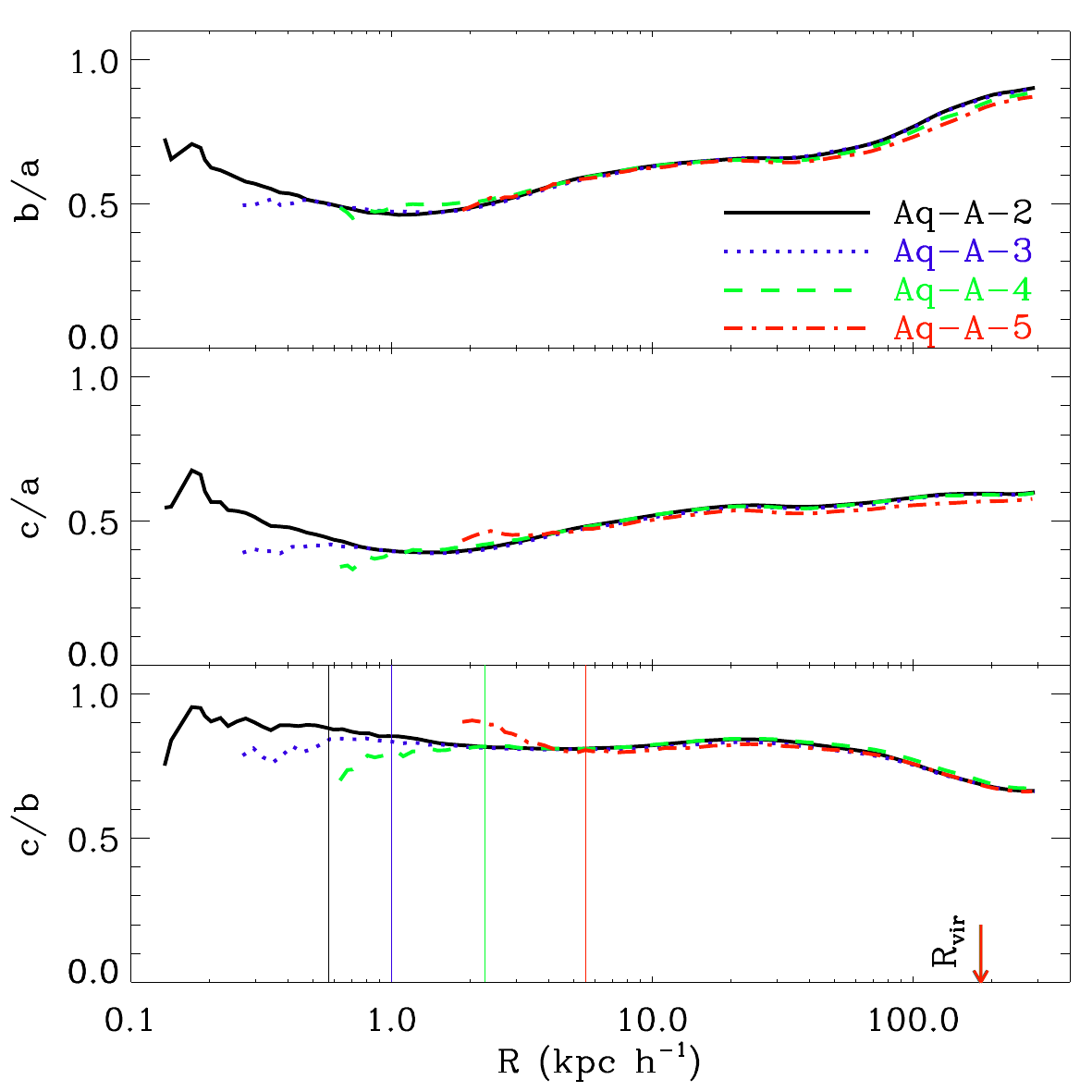}
  \caption{\label{axis-ratios-convergence} Convergence in the axis
    ratios for the \Aq{A} halo at four different numerical
    resolutions. The agreement is remarkable showing that the 
    method used provides reliable axis ratios down to the ``convergence
    radius'' $r_{\rm conv}$ (denoted by the vertical lines).}
\end{figure} 

\begin{figure}
  \includegraphics[width = 0.477\textwidth]{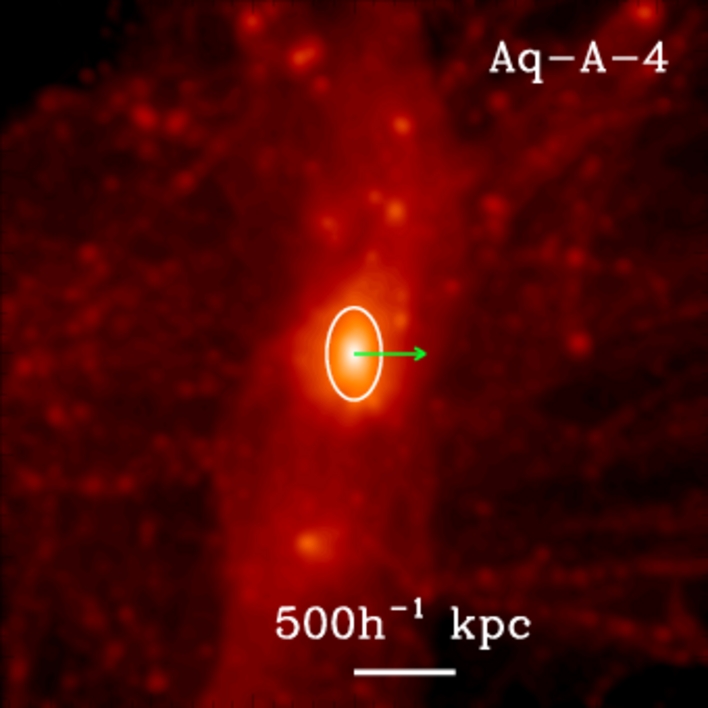} \\
  \includegraphics[width = 0.236\textwidth]{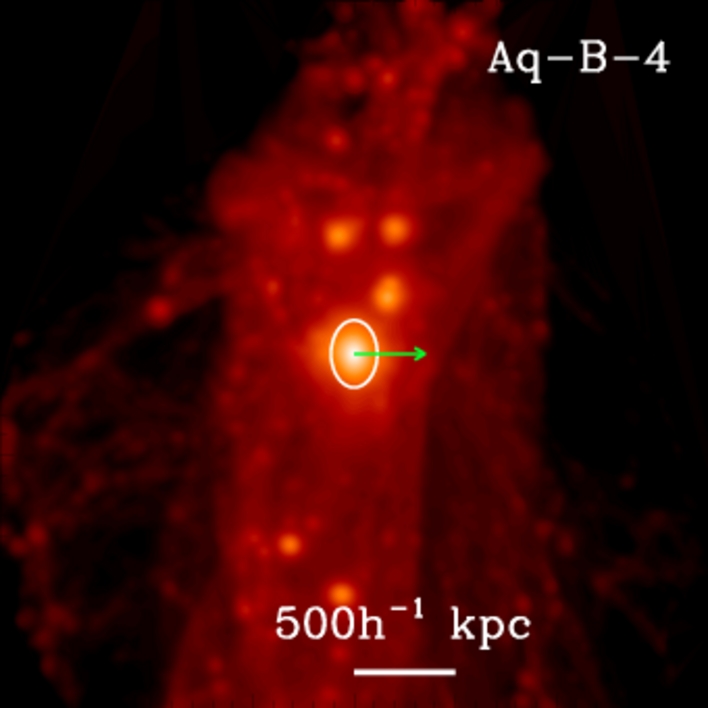}
  \includegraphics[width = 0.236\textwidth]{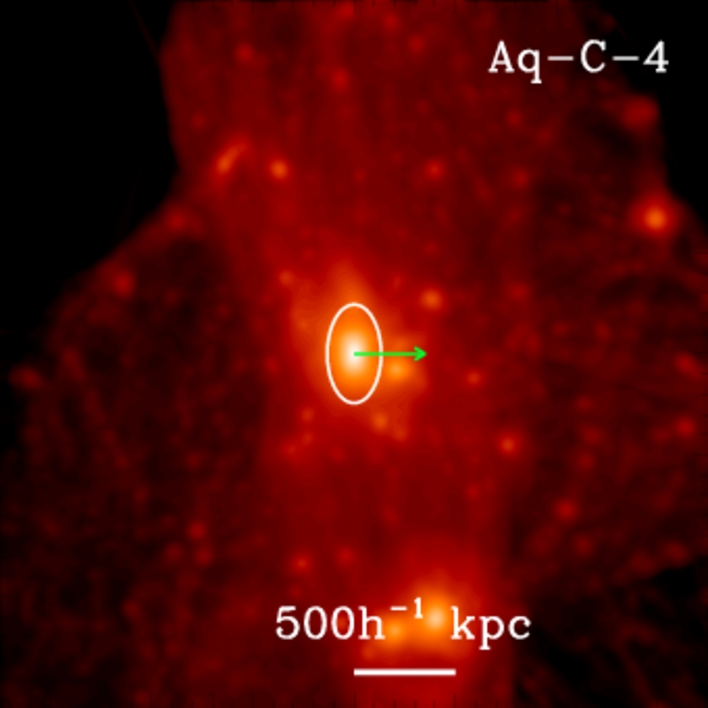}
  \includegraphics[width = 0.236\textwidth]{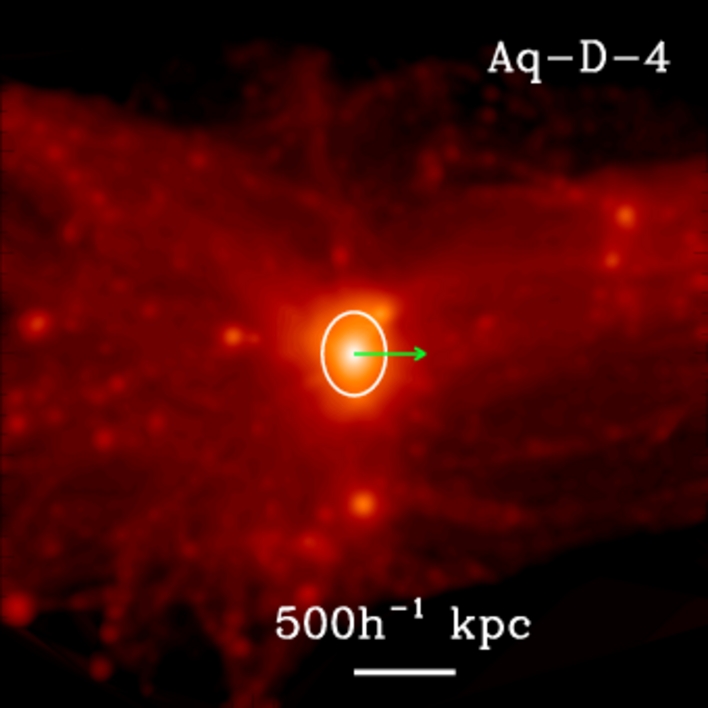}
  \includegraphics[width = 0.236\textwidth]{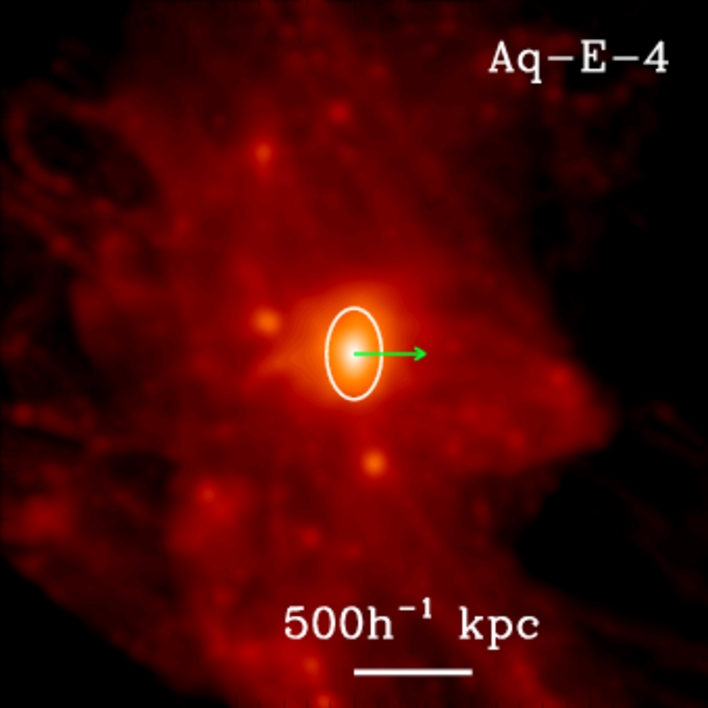}
  \caption{\label{present-day-density} Present day dark matter
  distribution of the Aquarius haloes and their surroundings. Colors
  are proportional to the logarithm of the squared dark matter density
  integrated along the line of sight. In each panel the depth of the
  projected image is $1.5h^{-1}$ Mpc. The virial ellipsoid is
  indicated in each panel with a solid white line and the green arrow
  shows the direction of the minor axis of the halo at the virial
  contour. The system is orientated such that the minor axis points
  horizontally and the major axis points vertically.}
\end{figure}


Various methods have been introduced in the literature to measure the
shapes of dark matter haloes. In Appendix \ref{sec:methods-appendix}
we describe a wide range of these methods in detail, and show that the
measured halo shapes agree reasonably well when applied to the same
object. Throughout this article we use the ``reduced'' inertia tensor
method as implemented recently by \cite{Allgood2006}. This tensor is
defined as

\begin{equation}
  I_{ij} = \sum_{\vect{x}_k \in V} \frac{x_k^{(i)}x_k^{(j)}}{d_k^2},
\end{equation}

\noindent where $d_k$ is a distance measure to the $k$-th particle and
$V$ is the set of particles of interest.  Assuming that dark matter
haloes can be represented by ellipsoids of axis lengths $a \ge b \ge
c$, the axis ratios $q=b/a$ and $s=c/a$ are the ratios of the
square-roots of the eigenvalues of $\vect{I}$, and the directions of
the principal axes are given by the corresponding
eigenvectors. Initially the set $V$ is given by all particles located
inside a sphere which is re-shaped iteratively using the eigenvalues
of $\vect{I}$. The distance measure used is $d_k^2 = x_k^2 + y_k^2/q^2
+ z_k^2/s^2$, where $q$ and $s$ are updated in each
iteration. Furthermore, in the figures below we have removed all bound
substructures contained in a halo using the \textsc{Subfind} algorithm
\citep{Springel2001}.  This alleviates the noise and artificial
tilting of the ellipsoids that is introduced by such
substructures. More details on our implementation can be found in the
Appendix \ref{sec:methods-appendix}.

We test the convergence of the shape measurements using the different
resolutions of the halo \Aq{A}, from level 5 to level 2. There is an
increase of a factor $\sim 230$ in the number of particles within the
virial radius and a factor $\sim 10$ reduction in gravitational
softening from \Aq{A-5} to
\Aq{A-2}. Fig.~\ref{axis-ratios-convergence} shows the axis ratios as
a function of $R$ for halo \Aq{A}, where $R$ is defined as the
geometrical mean of the axis lengths $R = (abc)^{1/3}$. This quantity
then provides a notion of distance to the centre of the ellipsoid with
the advantage of being volume invariant. Hereafter, we use $R$ when
referring to quantities measured using ellipsoidal contours, and $r$
for spherical contours. With this convention, $r_{\rm vir}$ is the
spherical radius enclosing an average overdensity of 200 times the
critical value, whilst $R_{\rm vir}$ refers to the ellipsoid enclosing
the same overdensity.

\begin{figure*}
  \centering \includegraphics[width =
  0.45\textwidth]{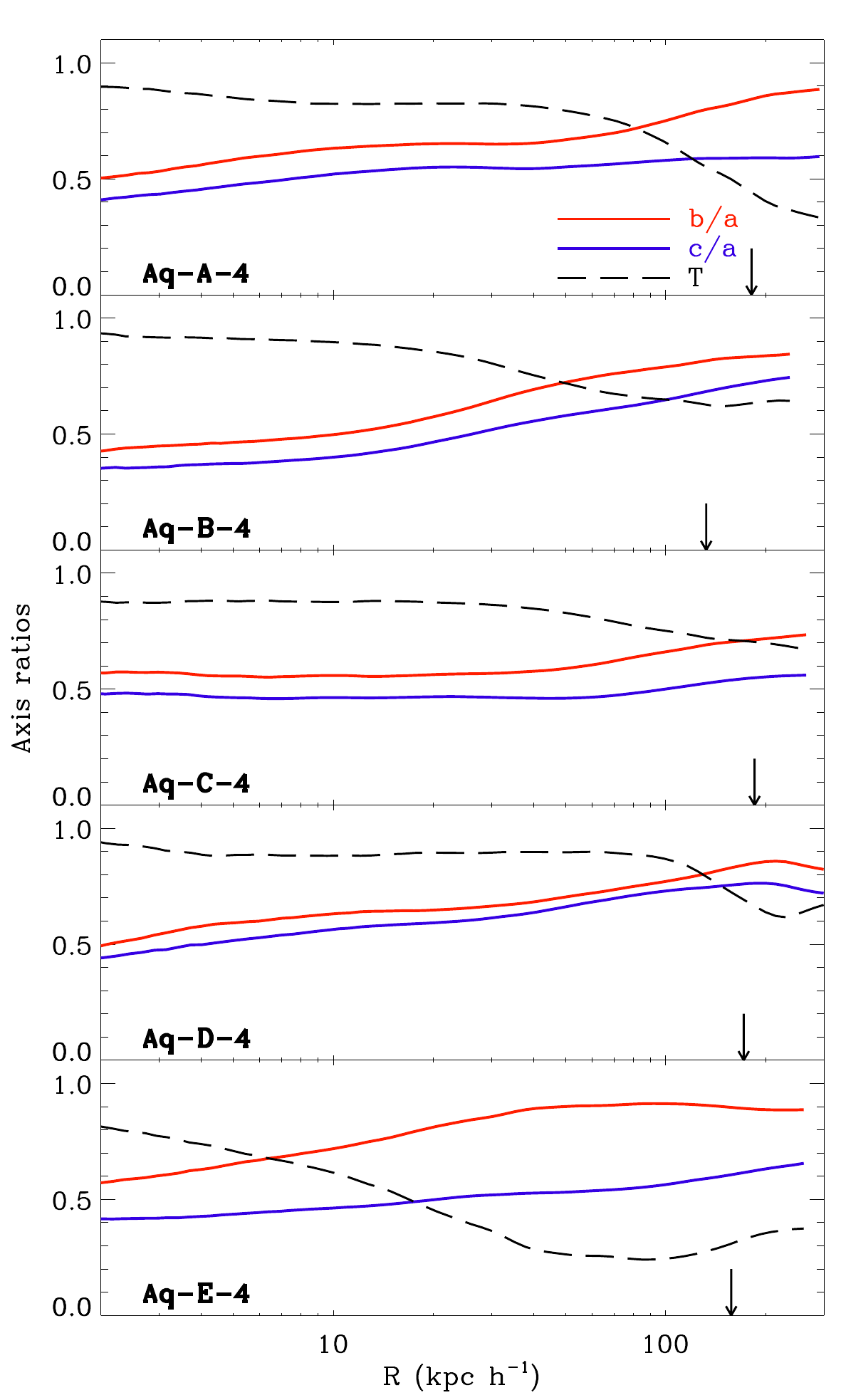}
  \centering \includegraphics[width =
  0.45\textwidth]{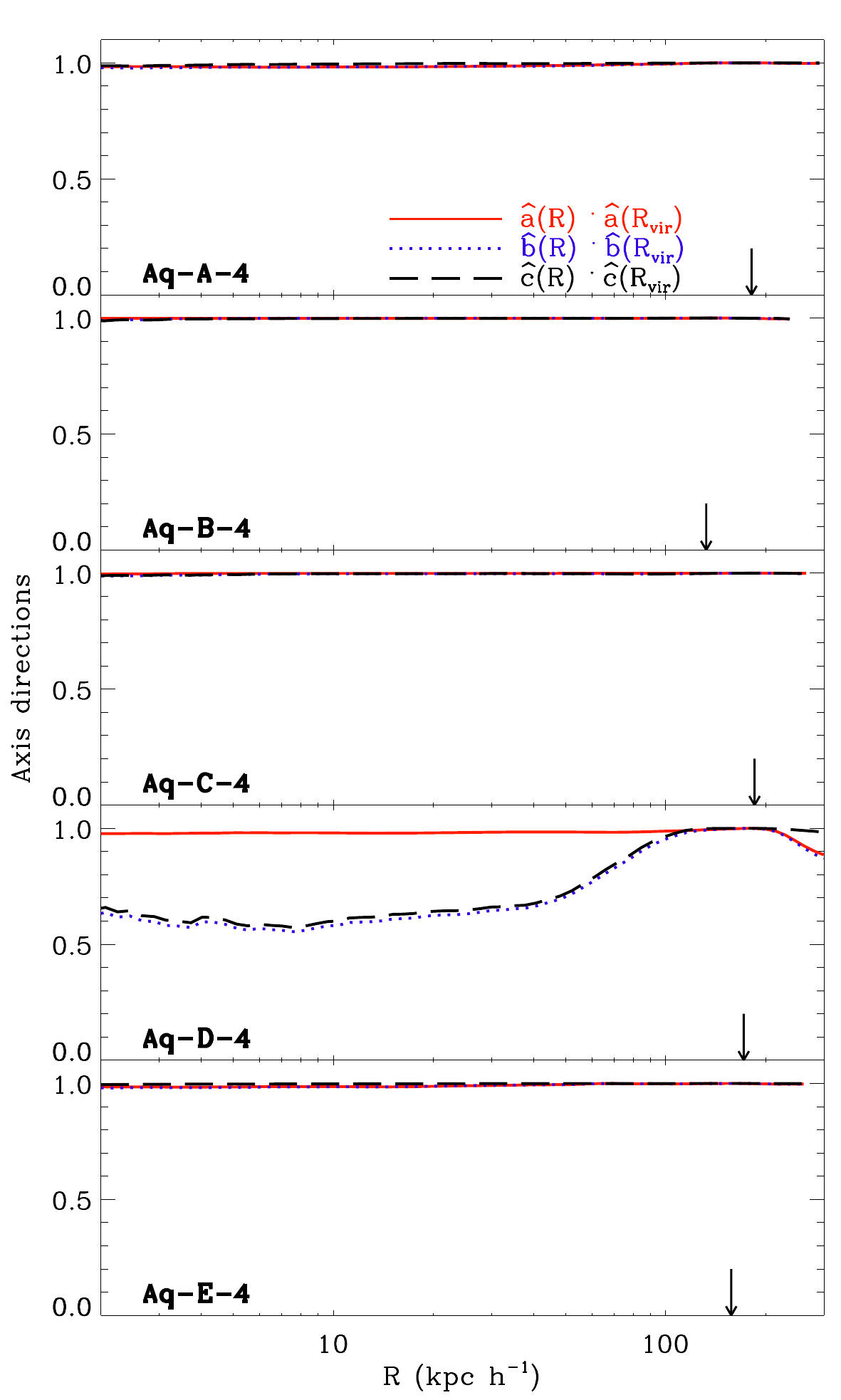}
  \caption{\label{axis-ratios-vs-R} Axis ratios (left) and directions
    (right) as a function of $R$ for each of the Aq-haloes. In the left
    panels the thick solid lines represent the axis ratios $b/a$ (red)
    and $c/a$ (blue), while the dashed curves are the triaxiality
    parameter $T$. In general haloes are more prolate in the inner
    parts and more oblate (and triaxial) in the outskirts. All
    concentric shells are also strongly aligned with the exception of
    halo \Aq{D} which exhibits a clear twisting close to the virial
    radius.}
\end{figure*} 

The agreement between the different resolutions in
Fig.~\ref{axis-ratios-convergence} is remarkable at all radii,
although small deviations are present in the inner regions where
resolution effects are expected to be important. The vertical lines in
the lower panel indicate the ``convergence radius'' $r_{\rm conv}$
\citep{Power2003,Navarro2010}. This radius is defined by setting
$\kappa(r_{\rm conv}) = t_{\rm relax}(r_{\rm conv})/t_{\rm
circ}(r_{\rm vir}) = 7$, where $t_{\rm relax}$ is the local relaxation
time and $t_{\rm circ}(r_{\rm vir})$ corresponds to the circular orbit
timescale at $r_{\rm vir}$. This choice ensures that circular velocity
profiles for $r>r_{\rm conv}$ deviate less than 2.5\% from the highest
resolution value. Fig.~\ref{axis-ratios-convergence} shows that
$r_{\rm conv}$ provides also a very good estimate for the smallest
radius at which the halo shape has converged. Although not explicitly
shown, we have tested that the orientation of the ellipsoids is a well
defined attribute independent of resolution for $r_{\rm conv}<R<R_{\rm
vir}$.

Our analysis suggests that convergence in halo shapes is achieved in
the level 4 from approximately $2h^{-1}$~kpc onwards (Table
\ref{aquarius-details-table}). We will therefore focus in what follows
on the study of the different Aquarius haloes \Aq{A} to \Aq{E} at the
level of resolution 4, since this yields a robust characterization of
the dark matter halo shapes and orientation, while keeping the
numerical cost constrained.


\section{The shape of dark matter haloes}
\label{sec:shape}

\subsection{Present day}
\label{subsec:shape-radius}

\begin{figure*}
  \centering
  \includegraphics[width = 0.23\textwidth]{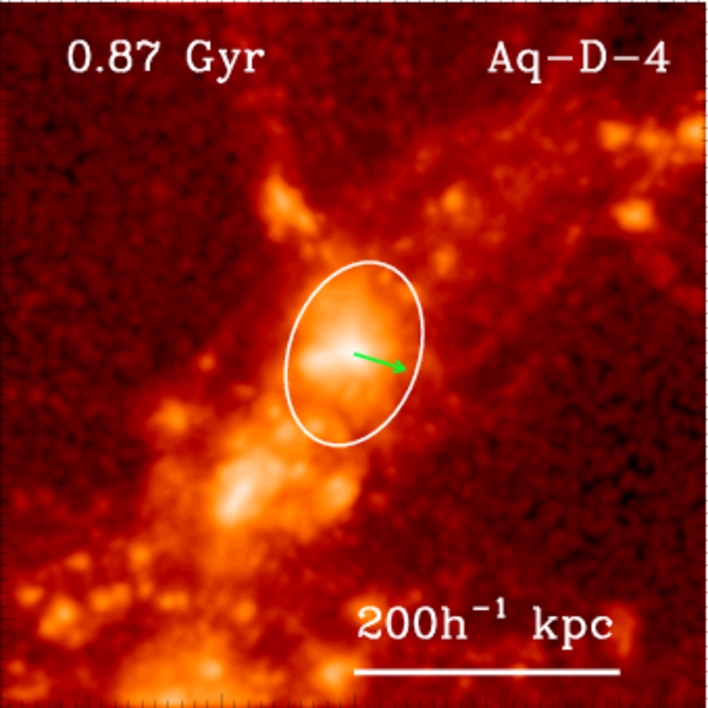}
  \includegraphics[width = 0.23\textwidth]{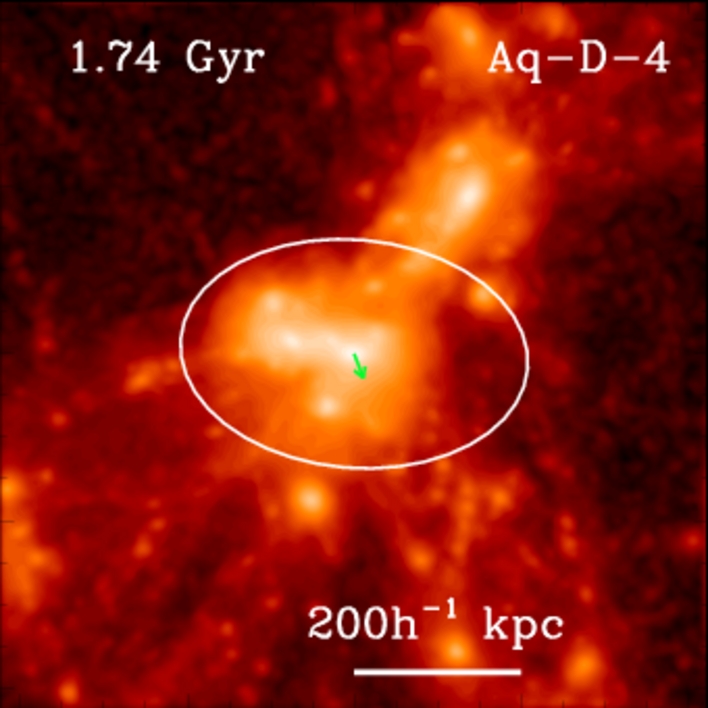}
  \includegraphics[width = 0.23\textwidth]{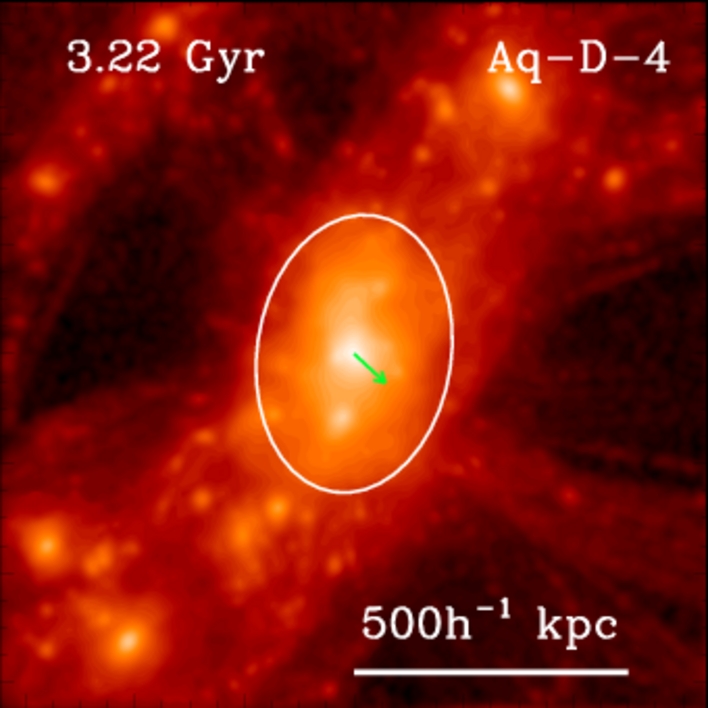}
  \includegraphics[width = 0.23\textwidth]{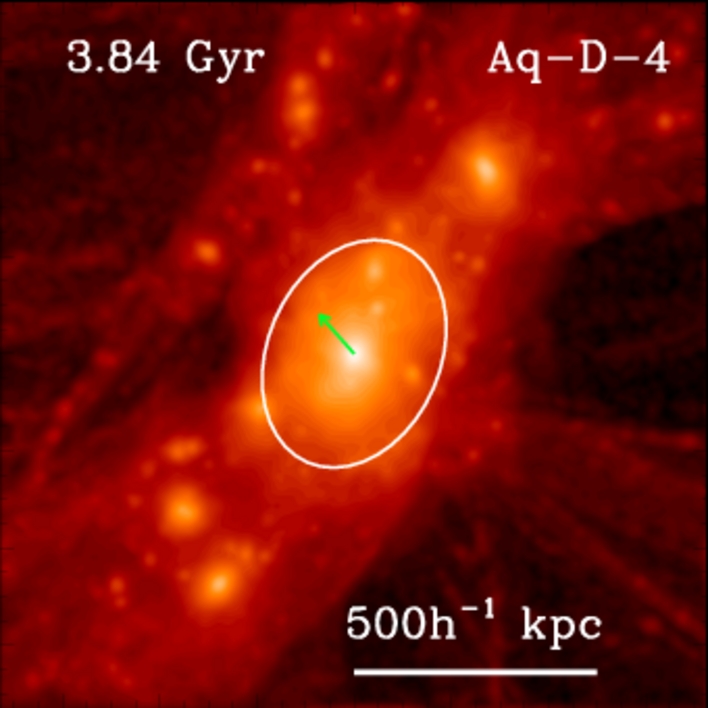}\\
  \includegraphics[width = 0.23\textwidth]{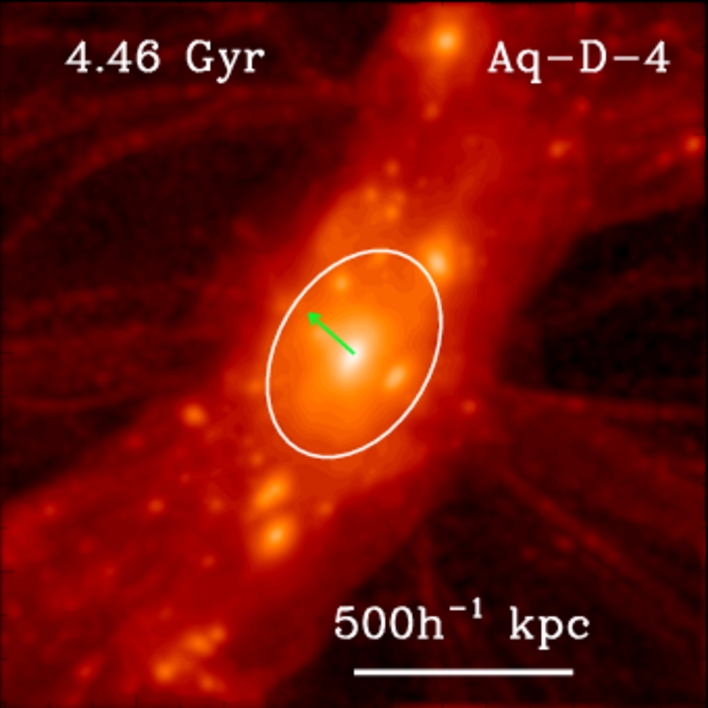}
  \includegraphics[width = 0.23\textwidth]{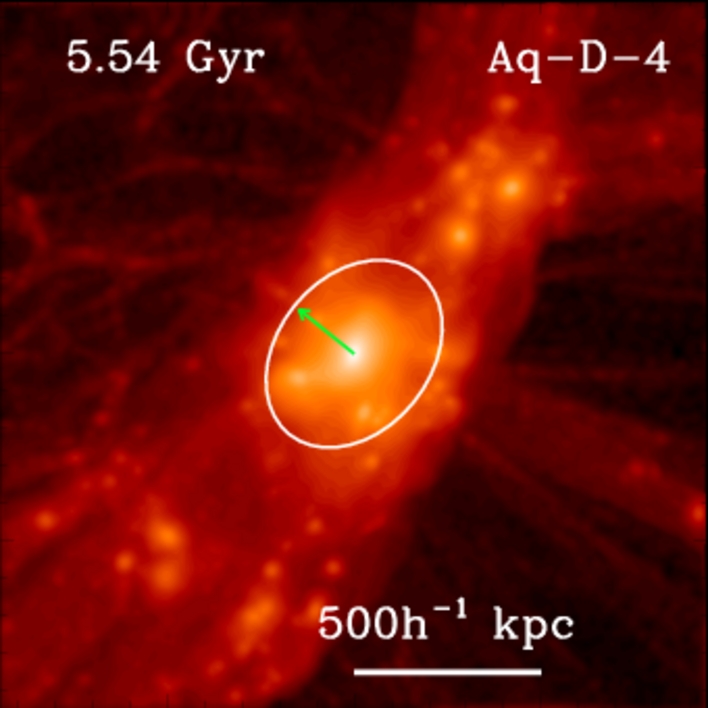}
  \includegraphics[width = 0.23\textwidth]{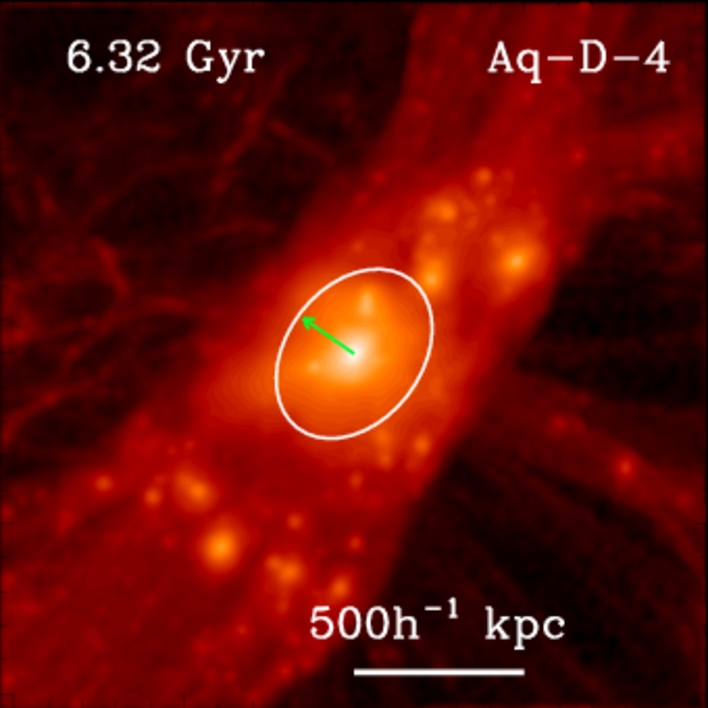}
  \includegraphics[width = 0.23\textwidth]{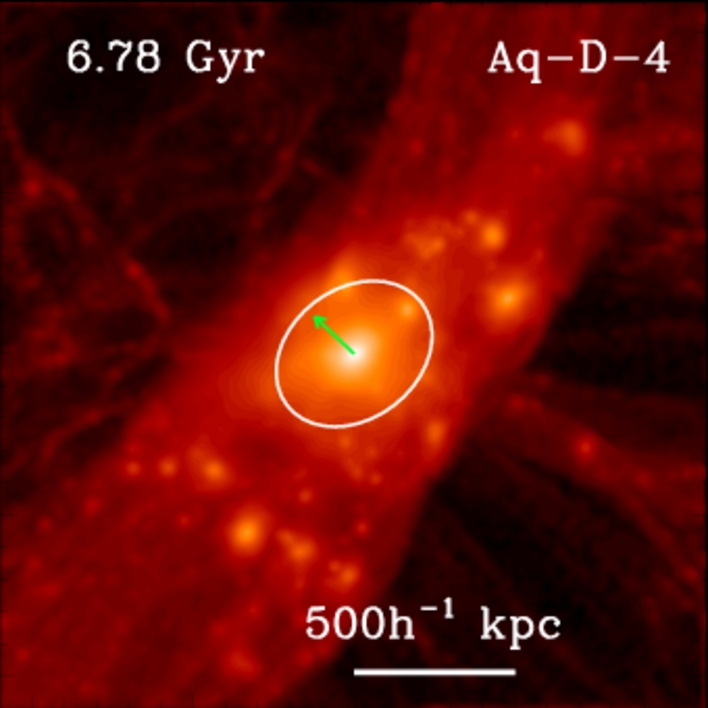}\\
  \includegraphics[width = 0.23\textwidth]{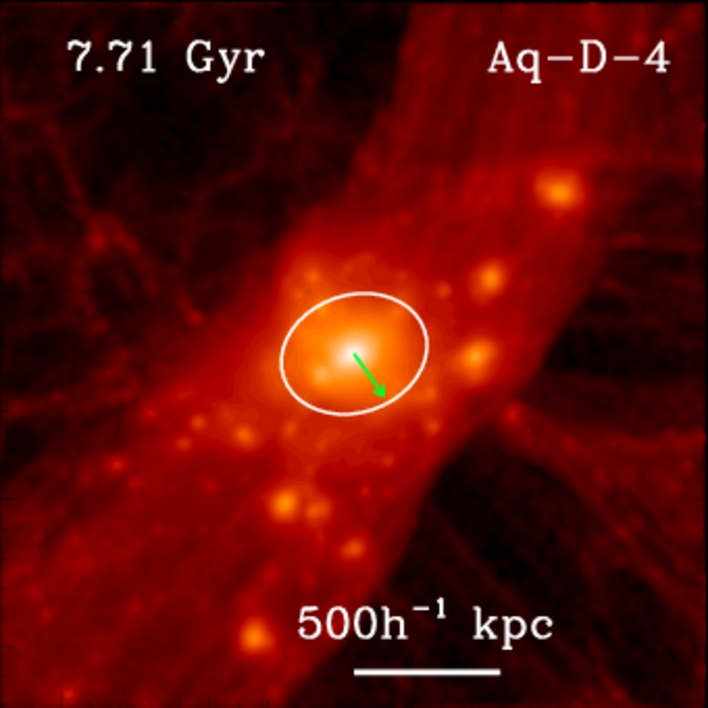}
  \includegraphics[width = 0.23\textwidth]{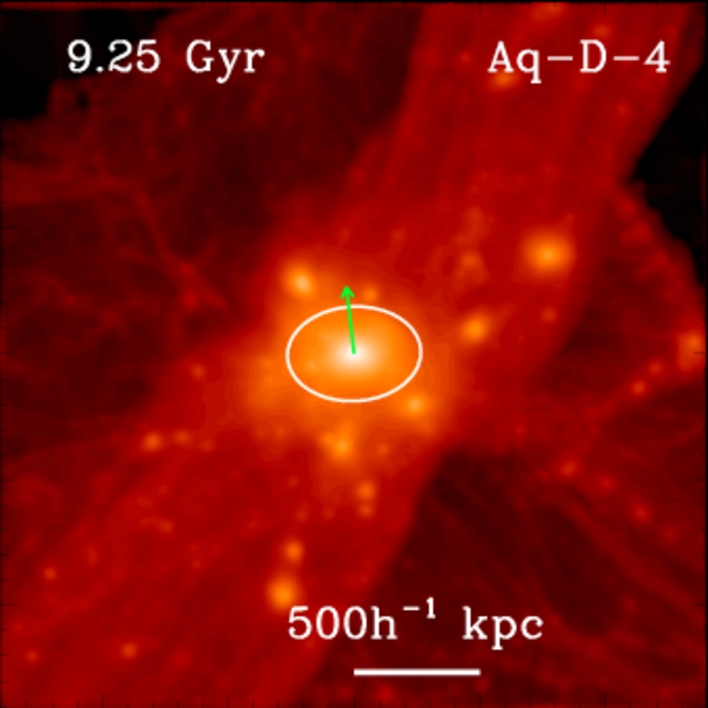}
  \includegraphics[width = 0.23\textwidth]{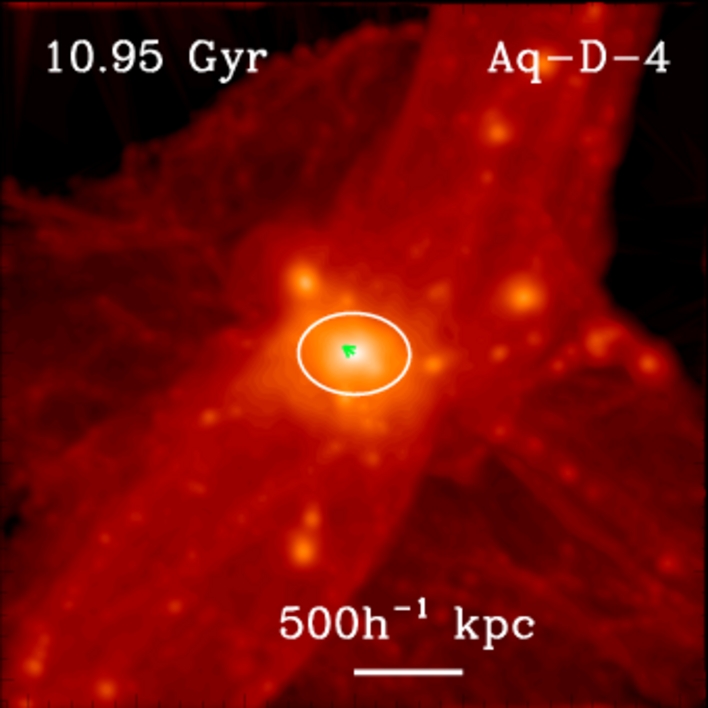}
  \includegraphics[width = 0.23\textwidth]{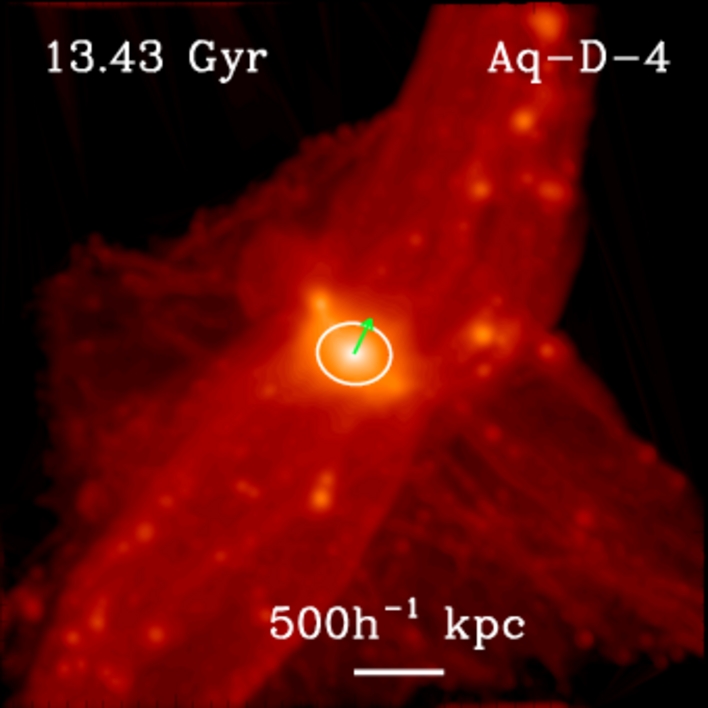}\\
  \caption{Time sequence of the formation of halo \Aq{D} from $t \sim
    1$ Gyr to the present day. Each box shows the dark matter density
    distribution around this halo together with the shape and
    orientation of the virial contour at the given time. As before,
    the green arrows indicate the projection of the minor axis onto
    the plane, with the length of the arrow being proportional to the
    minor axis length. The shape of this halo evolves continuously
    with time, showing stretching and tilting in response to the
    infall of matter. The halo is oriented along the filament up to $t
    \sim 7$ Gyr and then rotates and becomes perpendicular to it at
    the present day. This figure clearly illustrates the complexity
    involved in the build up of the present day dark matter halo's
    structure.}
  \label{movie-fig} 
\end{figure*}

Fig.~\ref{present-day-density} shows a snapshot view of the Aquarius
haloes and their surrounding environment. The color scale used is
proportional to the logarithm of the squared dark matter density
integrated along the line of sight, with a projection depth per panel
of $1.5h^{-1} \rm Mpc$. All images have been rotated according to the
orientation of the virial contours (indicated by the solid white
curve) in such a way that the major and minor axis define the vertical
and horizontal direction, respectively.

A careful look at the environment of each object reveals that the
Aquarius haloes are all embedded within a filamentary-like structure,
which is better defined in some cases than in others
(e.g. Fig.~\ref{present-day-density} \Aq{A} vs \Aq{E}), a result that
holds independently of the projection used. These filaments also seem
to contain the most massive substructures in the box and
interestingly, the minor axes of the virial contours tend to be
perpendicular to the direction defined by these.  This is in very good
agreement with the statistical findings of \citet{Bailin2005}, who
analyzed a sample of $\sim 4000$ haloes in a wide range of masses. One
relevant consequence of this type of configuration is that most of the
substructure will preferentially be accreted along the {\it major}
axis of the halo, a feature that has been used to explain the
preferential alignment of satellite galaxies with respect to their
hosts in observational studies of external galaxies \citep[e.g.][and
references therein]{Brainerd2005}, in groups
\citep[e.g.][]{Bingeli1982, Kitzbichler2003, Yang2006,
Faltenbacher2007, Godlowski2010, Paz2011} and also numerical
simulations \citep[e.g.][]{Knebe2004, Libeskind2005, Libeskind2007,
Kang2007, Sales2007, Bailin2008, Faltenbacher2008, Libeskind2011}.

The halo shapes and orientations as a function of radius are shown in
Fig.~\ref{axis-ratios-vs-R}. On the right panel, besides the axis
ratios, we also plot the triaxiality parameter $T$ defined as
\begin{equation}
  T = \frac{a^2-b^2}{a^2-c^2},
\end{equation}
which is unity for a perfect prolate distribution and zero in the
oblate case \citep{Franx1991, Warren1992}. This figure shows that in
the inner regions all dark matter haloes are more ``prolate'' with
$b/a \sim c/a \sim 0.4$--$0.6$ \citep{Hayashi2007}. On the other hand,
at large radii $b/a$ increases to $0.8$--$0.9$, and the mass
distribution becomes more oblate/triaxial.  At intermediate radii
haloes are typically triaxial, but notice that the radius at which the
prolate-to-oblate transition occurs (defined as the radius where
$T=0.5$) is different for each object, ranging from $\sim 20 h^{-1} $
kpc for \Aq{E} to $\sim 100 h^{-1}$ kpc in the case of \Aq{A}.

Regardless of the change in shape of the concentric ellipsoids, these
tend to remain well aligned throughout the halo. The right panel of
Fig.~\ref{axis-ratios-vs-R} shows the cosines of the relative angle
between the major, intermediate and minor axis of the ellipsoids at
different radii and at the virial contour. The alignment is
remarkable, with the exception of halo \Aq{D}, where the intermediate
and minor axis in the inner regions are rotated $\sim 60^\circ$ with
respect to their orientation at the virial radius, an issue that we
explore in the next section.


\subsection{Evolution}
\label{subsec:shape-time}

Fig.~\ref{movie-fig} shows a series of snapshots of halo \Aq{D} at
different stages of its evolution. The orientation of the coordinate
system is now kept fixed for all the snapshots. The corresponding time
is quoted in the upper left corner of each panel, the virial
ellipsoids are depicted by white ellipses while the projection of the
minor axes onto the plane are indicated by the green arrows. There are
several points worth highlighting:

\begin{itemize}
\item The shape and orientation of the halo seem to change throughout
  time (with the caveat that we are just seeing its evolution in {\it
  projection}).
\item Also evident from this figure is the filamentary structure that
  characterizes the surroundings of \Aq{D}. Notice that the filament
  where this halo is located at the present day was already in place
  at $t \lesssim 1$ Gyr, and fully dominates the environment from $t
  \sim 3$ Gyr onwards, with remarkable coherence in time and
  direction.
\item The relative orientation of halo \Aq{D} with respect to its
  environment is interesting since its minor axis is {\it parallel} to
  the direction of the filament at late times. This seems to be at
  odds with expectations from statistical studies of dark matter halo
  alignments \citep{Bailin2005, Faltenbacher2005, Aragon2007,
  Zhang2009} as well as in contradiction with the other haloes
  illustrated in Fig.~\ref{present-day-density} and the analysis
  presented in Section \ref{subsec:shape-radius}. A closer inspection
  of the history of this halo shows that the infall of material at $t
  \gtrsim 7.5$ Gyr occurs mostly along a secondary filament (only
  barely visible in this projection) whose direction is almost
  perpendicular to that of the most prominent and massive filament
  that dominates the surrounding large scale structure and is clearly
  seen in Fig.~\ref{movie-fig}.  This change of infall direction
  explains the apparent change in orientation of halo \Aq{D} at late
  times.
\item The relative size of a filament with respect to that of the halo
  increases with time: the filament's cross section is comparable to
  the virial radius at $t \lesssim 5$ Gyr, whereas the dark matter
  halo becomes smaller than and is fully embedded in the filament at
  later times. This has interesting consequences on the halo shape as
  we discuss further in Section \ref{sec:lss}. We have explicitly
  checked that this conclusion is not dominated by projection effects
  nor rendering of our images. We refer the reader to Appendix
  \ref{sec:filament_size}, where we introduce a physically motivated
  definition of a filament (based on the number of caustic-crossings a
  particle has experienced) and compare its size with that of the halo
  at each timestep.
\end{itemize}

\begin{figure}
  \centering \includegraphics[width =
  0.45\textwidth]{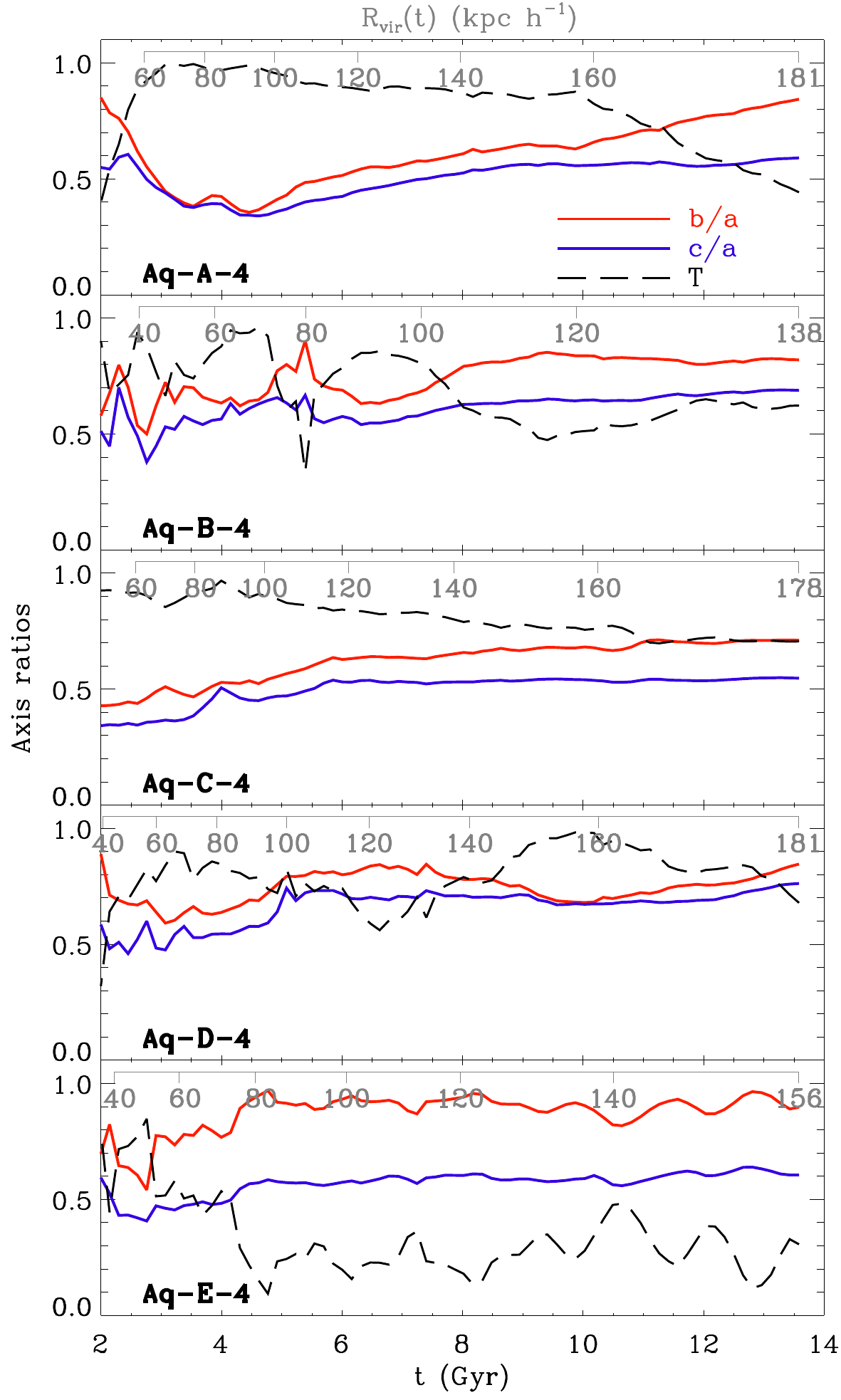}
  \caption{\label{axis-ratios-vs-time} Temporal evolution of the axis
    ratios and triaxiality parameter for the virial contour. The
    color-coding is the same as in Fig.~\ref{axis-ratios-vs-R}. A
    comparison to that figure shows that a correlation exists between
    the shape as a function of time, and the present-day shape as a
    function of radius. To aid this comparison we have added a second
    axis (gray), indicating the size of the virial contour $R_{\rm
    vir}$. This is derived for each halo by fitting: $R_{\rm vir}(t) =
    A (1 - {\rm e}^{(t-\tau)/(2 t_h)})$ to the measured evolution of
    the virial (ellipsoidal) radius.  Here $t_h$ is the time at which
    the virial mass reaches half its present value, and $A$ and $\tau$
    are free parameters.}
\end{figure}

In Fig.~\ref{axis-ratios-vs-time} we quantify the time evolution of
the axis ratios measured at the instantaneous virial ellipsoids for
haloes \Aq{A} to \Aq{E}. For reference, we have indicated with grey
labels the corresponding physical sizes at a given time. We will only
follow the evolution from $t=2$~Gyr onwards (which corresponds roughly
to the time at which the haloes have $\sim 20\%$ of their final mass),
which guarantees that the centre of mass as well as the shapes at the
virial radius are well defined at all times for the full sample of
haloes.

Fig.~\ref{axis-ratios-vs-time} shows that the shape of the virial
ellipsoids is not constant with time. In general, haloes seem to
evolve from a quite prolate configuration at early times towards more
triaxial shapes at the present day. In most of our haloes (with the
exception of \Aq{D}), this evolution in shape is driven by a larger
increase of the intermediate-to-major axis ratio $b/a$ than in
$c/a$. Except for this weak general trend, the evolution of the axis
ratios is quite disparate from halo to halo.

A qualitative comparison between the left panel of
Fig.~\ref{axis-ratios-vs-R} and \ref{axis-ratios-vs-time} suggests a
certain degree of resemblance between the overall evolution of the
virial ellipsoid's shape with time (as measured, for example, by the
triaxiality parameter) and the shapes measured as a function of radius
(Fig.~\ref{axis-ratios-vs-R}) for each of our haloes, despite their
intrinsic differences. This suggests that the dark matter haloes retain
certain {\it memory} of their configuration in the past, and that this
is imprinted on their present day structure. We note however that this
refers to the overall {\it trend} with radius/time and does not imply
that it is possible to recover the {\it exact} numerical value of the
axis ratios at a given time from their present day value at a given
radius.

\begin{figure*}
  \centering \includegraphics[width =
  0.99\textwidth]{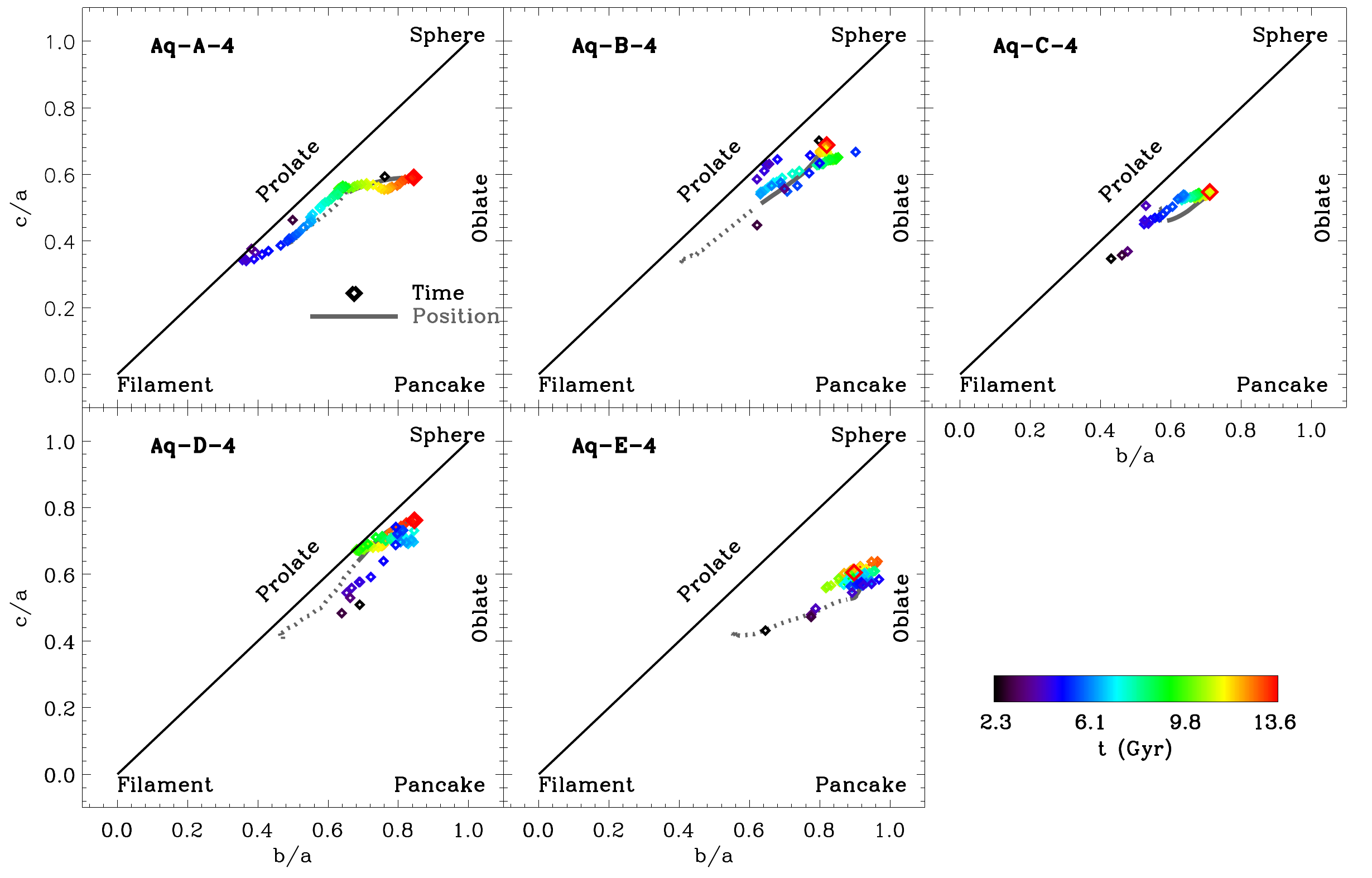}
  \caption{\label{axis-ratios-S-vs-Q} Axis ratios as a function of
    position (gray line) and time (diamonds). The colors indicate the
    time when the axis ratios at the virial contour have been
    measured. Notice that since we consider temporal evolution only
    after $t=2$ Gyr, we have differentiated the shape as a function of
    position (gray curve) using solid (dotted) lines for distances
    larger (smaller) than the virial radius of each halo at $t=2$ Gyr
    (see text for details).}
\end{figure*}

This analogy between radius/time is better seen in
Fig.~\ref{axis-ratios-S-vs-Q}, where we plot the present day axis
ratios $b/a$ vs\ $c/a$ at different radii (grey curve) as well as
those measured at the virial ellipsoids for different times (color
points). Because we sample from $t=2$~Gyr onwards, a dotted or a solid
line is used to indicate respectively, the shapes at radii smaller or
larger than the virial ellipsoid $R_{\rm vir}$ at this time. This
figure shows that haloes evolve away from the 1:1 line and therefore
tend towards less prolate shapes with time (and also radius), with the
clear exception of halo \Aq{D}. Moreover, the reasonable agreement
between the solid curve and the diamonds in this figure suggests that
the present day shape of the dark matter haloes at different radii
provides information about the evolution of the virial ellipticities
during their assembly history.

The analysis presented in this section, if generalized to Milky
Way-type haloes, is directly relevant to the modeling and
interpretation of observational constraints on the shape of the
Galactic potential. As discussed in the Introduction, the Sagittarius
stream has led to contradictory results when modeled in axisymmetric
dark matter haloes with constant axis ratios (in the potential)
\citep{Ibata2001,Helmi2004,Law2005, Johnston2005, Fellhauer2006}. We
have shown here that significant variations in the axis ratios as a
function of radius exist for all our five Aquarius haloes and those
variations are linked to the evolutionary history of each
object. Although this might add an extra degree of freedom to models
that attempt to constrain the Galactic potential, our results from
Fig.~\ref{axis-ratios-S-vs-Q} indicate that by doing so it may be
possible to retrieve the local conditions around the Milky Way's halo
throughout its assembly \citep{Banerjee2011}.

\section{Environment, Mass Accretion and Halo Shapes}
\label{sec:lss}

Dark matter haloes continue to grow with time, through a regular,
although not always steady, injection of matter. Halos will respond to
this new material and re-structure themselves onto a new equilibrium
configuration as part of the virialization process. The environment of
an object determines to some extent the way in which the material is
incorporated into the halo. For instance, thin filaments imply
accretion through well-defined preferential directions whereas a more
isotropic mode is expected when the halo is embedded in a larger
structure. Several correlations of halo shapes with environment have
been found so far in simulations \citep[e.g.][]{Avila2005,
Faltenbacher2005, Patiri2006, Basilakos2006, Maccio2007, Hahn2007b}
and observations \citep[e.g.][]{Pimblet2005, Godlowski2010,
Niederste2010, Paz2011}, albeit with significant scatter. In this
section we explore in detail the role played by the environment and
mass assembly histories in shaping the Aquarius haloes.

The angular distribution on the sky of the infalling matter can
provide useful information about the preferred directions and modes of
the ongoing accretion \citep{Tormen1997,Colberg1999, Aubert2004,
Aubert2007, Libeskind2011}.  For instance, whereas isotropic accretion
would lead to a uniform signal on the sky, the presence of a thin
filament will give rise to a bi-modal distribution of points in two
opposite ($180^\circ$ apart) directions. This can be quantified
further by means of a multipole expansion of the infalling particles
on the sky at a given time.  We measure the power spectrum for the
mode $l$ as,

\begin{equation}\label{power-spect-eq}
 C_l = \frac{1}{4\pi} \frac{1}{2l + 1}\sum_{m=-l}^l
 |\tilde{a}_l^m|^2,
\end{equation}

\noindent where the expansion coefficients are,

\begin{equation}\label{power-spect-coeff-eq}
 \tilde{a}_l^m = \frac{m}{4\pi r_{\rm vir}^2}\sum_{k=1}^N (Y_l^m(\Omega_k))^*,
\end{equation}

\noindent with $\Omega_k$ being the angular position of the $k$-th
particle crossing the virial radius $r_{\mathrm{vir}}$ with negative
radial velocity at any given time. The asterisk on the right hand side
of Eq.~\eqref{power-spect-coeff-eq} indicates the complex conjugate of
the term within parenthesis.

In the scheme introduced above, the $l=0$ term (monopole) is a
constant equal to unity and used for the overall normalization of the
expansion.  Notice that although this choice is arbitrary, the {\it
relative} relevance of the monopole with respect to all the other
modes, $C_0/\Sigma_l C_l$, is an indication of how isotropic the
distribution is; i.e. a perfectly isotropic infall would correspond to
all the power in the $l=0$ term. On the other hand, a significant
contribution of the $l = 2$ (or quadrupolar moment) term arises when
the accretion occurs through a well-defined direction in space, i.e. a
filament. Similarly, accretion corresponding to more than one
preferential direction will shift the power away from $l=2$ and
towards higher moments. Notice that a point mass on the sky will
excite, by definition, a wide range of modes with similar power; just
like the Fourier transform of a Dirac-delta function has constant
power for all modes in the frequency space. We therefore expect single
satellite infall events to excite higher moments compared to the
smooth accretion. In the limit of a satellite that occupies a large
area of the sky, the configuration will then resemble a dipole and the
power spectrum will exhibit high power in the $l=1$ mode.

In practice, most of the information is encoded in the terms $l \leq
2$ \citep[e.g. ][]{Quinn1992,Eisenstein1995,Aubert2004}. We have
checked in our experiments that higher moments than quadrupolar
contribute always less than $\sim 15 \%$ of all power at any given
time once substructures have been removed. We therefore focus our
analysis mostly on the $l=0$ and $l=2$ moments since they provide most
of the information that allow to characterize the mass accretion onto
dark matter haloes.

\begin{figure}
 \centering \includegraphics[width =
 0.45\textwidth]{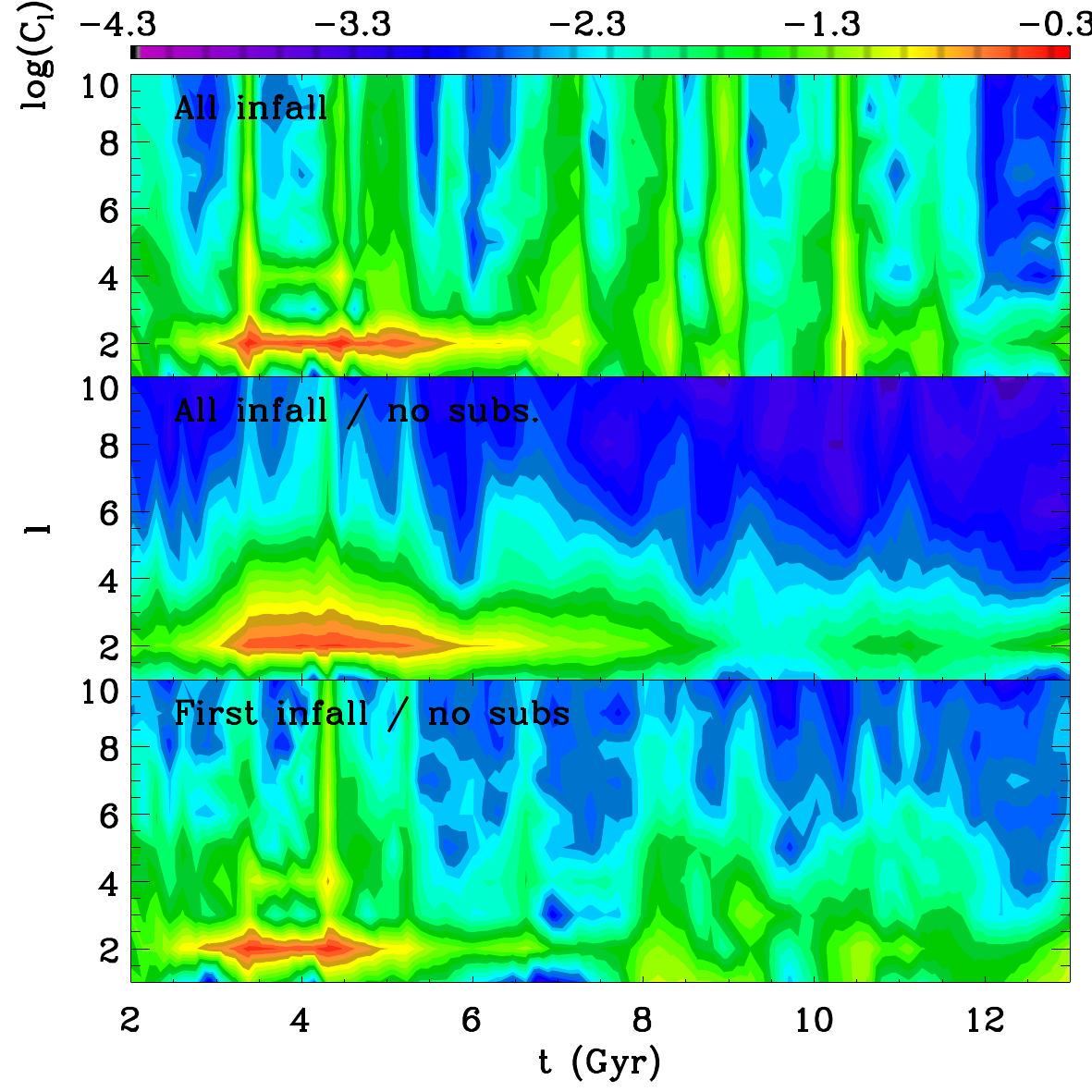}
 \caption{\label{power-spect-all-Aq-A-4-fig} Multipole expansion of
   the infalling material ($v_r<0$) in the region $1.0\leq r/r_{\rm
   vir} \leq 1.2$ as a function of time for halo \Aq{A-4}. The upper
   and middle panels show the results with and without the
   contribution from subhaloes, respectively. In the bottom panel only
   those particles that are on their {\it first infall} have been
   considered.}
\end{figure}

Fig.~\ref{power-spect-all-Aq-A-4-fig} shows this multipole
decomposition introduced in Eq.~\eqref{power-spect-eq} of the mass
infalling onto \Aq{A-4} as a function of time. At each output time, we
select particles with negative radial velocity (infalling), $v_r<0$,
that are in the spherical shell $1.0 \leq r/r_{\rm vir} \leq 1.2$ and
compute the corresponding $C_l$ of the distribution\footnote{We have
tested that the qualitative behavior shown in this figure and the
relative relevance of each mode with respect to the whole spectrum
does not depend on the particular shell that is analyzed for $1<
r/r_{\rm vir} <2$ for all of our haloes. }. The upper and the middle
panels of this figure show the $C_l$ power spectra obtained
respectively, by including and by removing particles associated with
substructures as identified by \textsc{Subfind}.  Both distributions
are quite similar, although as expected from the discussion in the
previous paragraph, satellite accretion excites in general higher
modes that last a very short timescale, and are visible as clear
``spikes'' in the upper panel of the figure.

The virialized regions of galaxy-sized objects can extend well beyond
their formal $r_{\rm vir}$ \citep[e.g. ][]{Cuesta2008}, introducing a
signal in the power spectrum that is driven by the halo's intrinsic
shape rather than by the surrounding infall pattern.  In order to
avoid confusion with the material already in place and in equilibrium
in the outskirts of the halo, we selected also the subsample of
particles within $1.0 \leq r/r_{\rm vir} \leq 1.2$ that are infalling
for the {\it first time} onto the halo (in practice we do this by
requiring that a particle in a given output has never been within the
virial radius of the main object at any previous time step). The
corresponding power spectrum (after removing the subhaloes'
contribution) is shown in the bottom panel of
Fig.~\ref{power-spect-all-Aq-A-4-fig}. This distribution agrees well
with those shown in the other panels of this figure, although the
features appear noisier due to the smaller number of particles being
considered.

\begin{figure*}
 \centering \includegraphics[width =
 0.45\textwidth]{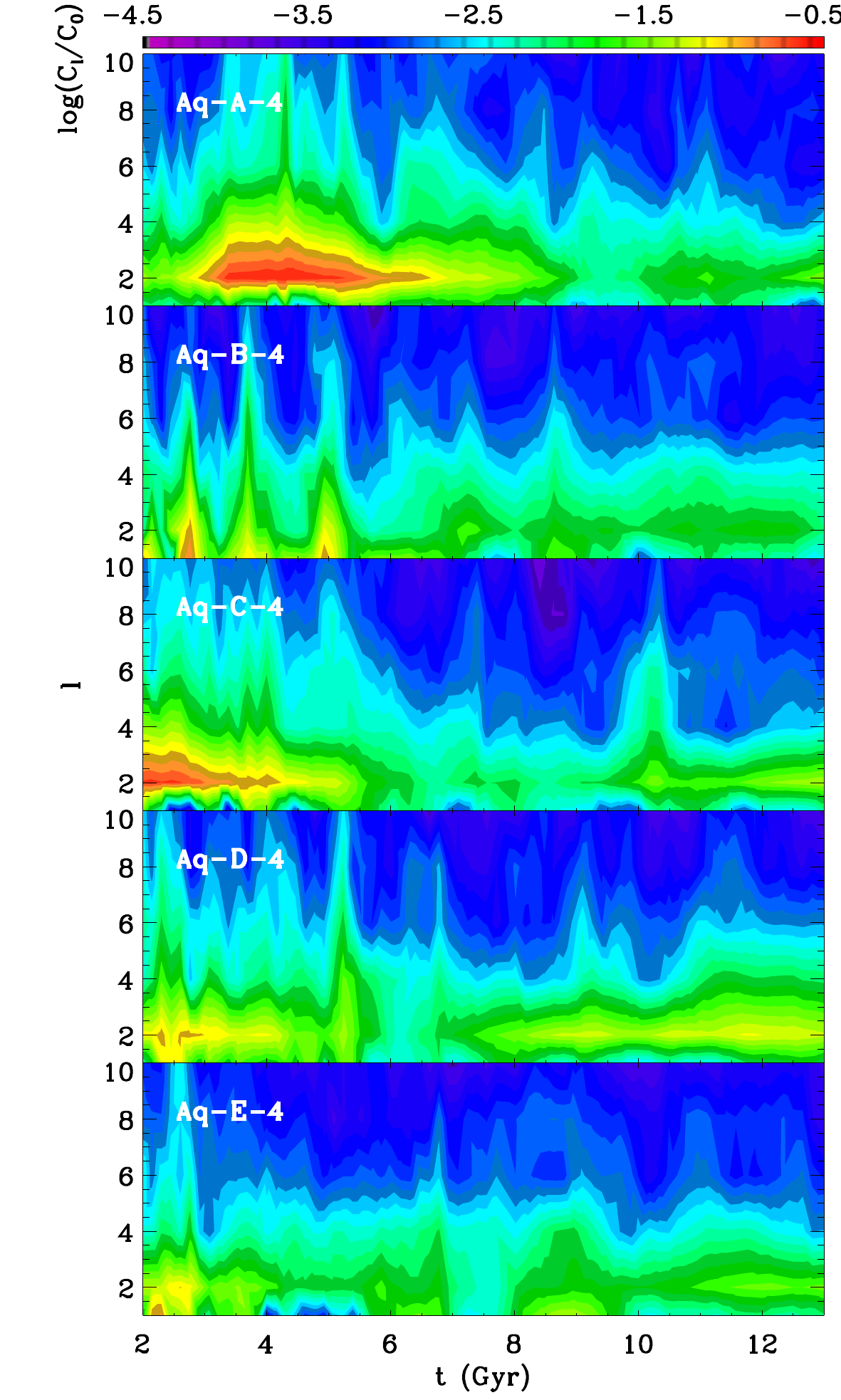}
 \centering \includegraphics[width =
 0.45\textwidth]{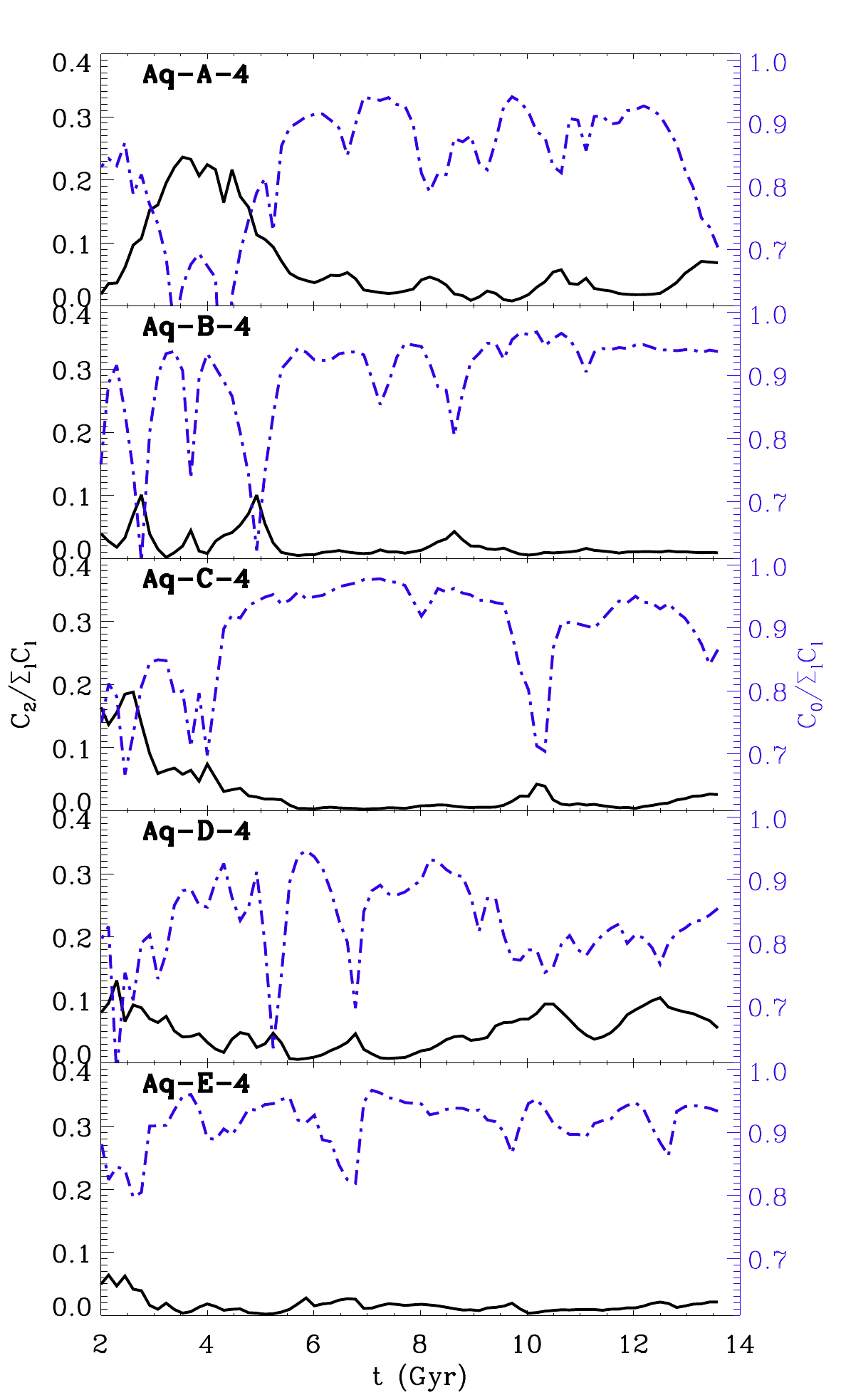}
 \caption{{\it Left:} Multipole expansion of the infalling material in
   the region $1.0\leq r/r_{\rm vir} \leq 1.2$ as a function of time
   once the contribution from subhaloes has been removed for the five
   Aquarius haloes. {\it Right:} Relative contribution of the $l=0$
   (blue dot-dashed) and $l=2$ (black solid) modes to the total power
   spectrum as a function of time for particles on their {\it first
   infall} onto each halo. While $C_2/\Sigma C_l$ provides information
   about the material infalling along a filament, large contributions
   from the monopolar term, $C_0/\Sigma C_l$, imply that the accretion
   is isotropic. Notice the good correlation between the time
   intervals with a clear signature of mass accretion through
   filaments (high power in $l=2$ mode) and the more prolate shape in
   Fig.~\ref{axis-ratios-vs-time}.}
\label{power-spect-modes} 
\end{figure*}

The left panel of Fig.~\ref{power-spect-modes} shows the power
spectrum of all infalling material for all five Aquarius haloes after
the contribution from subhaloes has been removed. The residual
``spikes'' visible in this figure correspond to the matter that is
associated to infalling substructures but that is rather loose and,
consequently, has not been assigned to a particular subhalo by the
halo finder.



The right panel of Fig.~\ref{power-spect-modes} shows the relative
contribution of the $l=2$ (solid black curve) and $l=0$ (blue
dot-dashed) moments to the total power spectrum as a function of
time\footnote{Note that we are not explicitly showing the $l=0$ term
in Fig.~\ref{power-spect-modes} because it is, by definition, set to
unity in our formalism. However, the ``relative'' importance of the
monopole with respect to the contribution of all other moments is a
well defined quantity, that we analyze in more detail on the right
panel of the same figure.}.  These have been computed using only the
subset of particles on their first infall (although these curves do
not change significantly when all infalling particles are considered
instead). Large $C_2$ values are associated with the presence of net
filamentary accretion.  Fig.~\ref{power-spect-modes} shows that this
condition is typically found at early times in our haloes, with the
exception of \Aq{B}, which shows no clear sign of smooth accretion
through a filament at any time. The halo \Aq{D} shows also a
peculiarly high power in the $l=2$ mode at late times, whereas for
most of the objects the relevance of $C_2$ remains approximately flat
(and negligible) in the last few Gyr (a feature which is also evident
in the left panel of Fig.~\ref{power-spect-modes}).

The monopole term largely dominates the accretion at later times (blue
dot-dashed curve in Fig.~\ref{power-spect-modes}) in all
haloes. Exceptions are haloes \Aq{A} and \Aq{D}, which show a decline
in $C_0/\Sigma\, C_l$ beyond $t\sim12$ Gyr and $t \sim 10$ Gyr
respectively. In the case of \Aq{A} this is related to an increase in
power of $l \ge 2$ modes, while for \Aq{D} this is driven only by the
$l=2$ moment, as mentioned in the previous paragraph.  As we show in
Appendix \ref{sec:filament_size}, this more isotropic infall is a direct consequence of
the increase in the relative size of the filaments feeding the dark
matter haloes: infalling particles cover wider angles on the sky which
leads to larger $C_0$ contributions with negligible $l=2$ component.
Notice that halo \Aq{E} is, in some sense, the most extreme object,
with a monopolar term that dominates the power spectrum of infall
material during almost all its entire history ($C_0/\Sigma_l C_l >
0.9$ for $t \sim 3$ Gyr onwards).

A careful comparison of Fig.~\ref{power-spect-modes} and
Fig.~\ref{axis-ratios-vs-time} reveals a good correlation between the
infall of material through filaments (large $C_2$) and the shape of
the haloes at a given time: high $l=2$ moments are associated with
virial contours that turn prolate (e.g. \Aq{A} for $t< 6$ Gyr, \Aq{C}
for $t< 4$ Gyr, \Aq{D} for $t>8$ Gyr). In particular, this multipole
decomposition shows a filamentary accretion mode for halo \Aq{D} that
is present at $t \geq 10$ Gyr, helping to explain why its axis ratio
$b/c$ remains close to unity at quite late times. When the accretion
is more isotropic (i.e. $C_0/\Sigma_l C_l \simeq 0.9$ ), the haloes
become most nearly oblate. This explains why halo \Aq{E} is the more
oblate in our sample, with $b/a \sim 0.9$ from $t\sim 3.5$ Gyr
onwards, and also why it has this shape at smaller radii, compared for
example to halo \Aq{C}.

We found that {\it the injection of material that occurs along
filaments leads to a more prolate halo shape, an effect that is
naturally enhanced at early times due to the relatively smaller size
of the filaments with respect to the dark matter halo}
\citep{Avila2005, Gottlober2006}. On the other hand, the more
isotropic mass accretion that characterizes later phases of halo
assembly for $10^{12}$M$_\odot$ Milky Way-like objects, yields more
oblate/triaxial geometries. When combined with the {\it memory} effect
alluded to in Section \ref{subsec:shape-time}, we find that prolate
shapes are naturally expected to be set in the earliest collapsed
regions (small radii) of $\sim 10^{12}$M$_\odot$ haloes, in good
agreement with previous work \citep{Bailin2005, Hayashi2007}. The
filamentary structure typically found at early times thus seems to be
responsible for the general trend of our Milky Way-like haloes to be
more prolate in their inner regions.



\section{Conclusions}
\label{sec:conclusions}

In this paper we analyzed the shape of five Milky Way-like dark matter
haloes selected from the Aquarius $N$-body simulations.  We compared
the performance of several methods proposed in the literature to
measure the shapes of haloes and found good agreement between all
techniques, especially in the inner regions where substructures play
only a minor role. Using an implementation of the normalized inertia
tensor algorithm described in \citet{Allgood2006}, we have found
excellent convergence between the several resolution levels of
Aquarius, where the shapes can be robustly measured down to the
convergence radius.

We find that mass assembly and environment are both responsible for
setting the shapes of dark matter haloes. The early evolutionary phases
of $10^{12}$M$_\odot$ Milky Way-like haloes are characterized by the
accretion of matter through narrow filaments. In these circumstances
the haloes --as measured by their virial contours-- are prolate and
their minor axes tend to point perpendicular to the infall (filament)
direction. Nonetheless, temporary tilting of the virial ellipsoids may
occur when mass is accreted from a different direction. The latter is
the case for just one of our haloes located in a well defined filament
at redshift $z=0$.

On the other hand, at later times the cross-section of the filaments
becomes larger than the typical size of Milky Way-like haloes and as a
result, accretion turns more isotropic and the objects evolve into a
more oblate/triaxial configuration. This transition does not occur at
the same time for all haloes in the explored mass range but is strongly
determined by their individual history of mass assembly and their
surrounding environment.

The geometrical properties of haloes at different epochs are not lost:
haloes retain memory of their structure at earlier times. This memory
is imprinted in their present-day shape trends with radius, which
change from typically prolate in the inner (earlier collapsed) regions
to a triaxial in the outskirts (corresponding to the shells that have
collapsed last and are now at the virial radius). These results are in
excellent agreement with previous findings \citep{Bailin2005,
Hayashi2007}.

A corollary of our results is that the strong link between halo
properties and assembly history, which can show large variations from
halo to halo, will make any {\it instantaneous} correlation between
haloes shape or orientation and mass or environment rather weak,
explaining in part the relatively large scatter in such trends found
in earlier studies \citep[e.g. ][]{Bailin2005,Bett2007}.

It is important to stress that we have neglected the effects induced
by the presence of baryons on the dark matter halo shapes
\citep{Kazantzidis2004, Bailin2005b, Gustafsson2006, Debattista2008,
Pedrosa2009, Lau2010, Valluri2010}. Therefore our results may only be
directly applicable to dark matter dominated objects such as low
surface brightness galaxies.  However, the work of \citet{Tissera2009,
Abadi2010, Bett2010} shows that even though the halo axis ratios
increase when a disc is formed (i.e.\ they become rounder), the trends
with radius appear to be preserved.

With the caveat of the neglected baryonic effects and the relatively
low number of objects studied, our findings may be directly relevant
to the modeling of stellar streams used to determine the gravitational
potential of the Milky Way. For example, the inconsistencies found
between the constraints imposed by the positions and by the kinematics
of stars in the Sagittarius stream could be indicative of a change in
the shape of the Milky Way halo with radius \citep[although
see][]{Law2009,Law2010}. Such a change could in principle be measured
by using stellar streams which are on different orbits in the Galactic
halo. Furthermore, it may even be possible to employ these to
determine the growth history and early environment of the Milky Way. A
validation of these ideas, however, is bound to a proper evaluation of
the effect of the baryonic matter on our findings, as well as to the
study of larger statistical samples, both issues that we plan to
address in the near future.

\section*{Acknowledgments}
LVS and AH gratefully acknowledge financial support from NWO and from
the European Research Council under ERC-Starting Grant
GALACTICA-240271. CSF acknowledges a Royal Society Wolfson Research
Merit Award. This work was supported in part by an STFC rolling grant
to the Institute for Computational Cosmology at Durham. We thank the
anonymous referee for useful suggestions and comments.

\bsp
\label{lastpage}

\bibliographystyle{mn2e}
\bibliography{refs}

\appendix
\section{The shape of dark matter haloes: methods}
\label{sec:methods-appendix}

In this section we provide a review of the various methods that have
been proposed in the literature to measure dark matter halo shapes and
compare the results when applied to the same object. The methods most
commonly used are the diagonalization of the inertia tensor and the
characterization with ellipsoids of either the interpolated density
field or the underlying gravitational potential
\citep{Warnick2008}. We introduce as well an additional scheme, which
incorporates the advantages of different pre-existing methods, and
where halo shapes are determined by fitting ellipsoids to the 3D
iso-density surfaces.

\begin{figure*}
  \centering \includegraphics[width =
  0.99\textwidth]{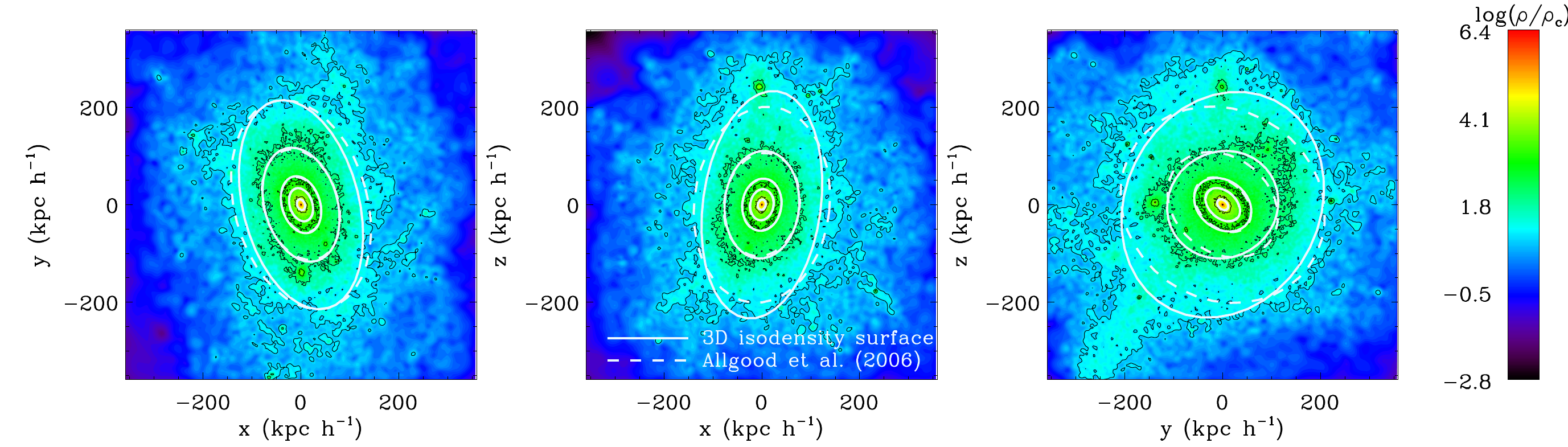}
  \caption{\label{density-contours} Dark matter density in thin slices
  centred on the $z=0$ (left) $y=0$ (centre) and $x=0$ (right)
  planes. The black lines represent the isodensity contours. The solid
  white ellipses correspond to our new density-based method, while the
  dashed ones are for the inertia tensor method described by
  \citet{Allgood2006}. In the inner regions both methods agree very
  well and follow the isodensity surface, however closer to the virial
  radius (the external most contour) the inertia tensor-based scheme
  predicts somewhat rounder shapes than the actual density
  distribution.}
\end{figure*} 

\subsection{Inertia tensor}

One of the drawbacks of methods based on the determination of the
inertia tensor is its quadratic relation with distance, which assigns
the largest weight to particles residing far away from the centre, an
issue complicated further by the presence of substructures in the
outskirts of the dark matter haloes.

The bias introduced by this distance weighting can be alleviated by
normalizing each coordinate with some measure of distance
\citep{Gerhard1983, Dubinski1991}. In this case the ``reduced''
inertia tensor

\begin{equation}
  I_{ij} = \sum_{\vect{x}_k \in V} \frac{x_k^{(i)}x_k^{(j)}}{d_k^2},
\end{equation}

\noindent where $d_k$ is a distance measure to the $k$-th particle and
$V$ is a set of particles' positions. Assuming that dark matter haloes
can be represented by ellipsoids, the axis ratios are the ratios of
the square-roots of the eigenvalues of $\vect{I}$, and the directions
of the principal axes are given by the corresponding eigenvectors. To
determine the axis lengths (e.g. $b$ and $c$) however requires
knowledge of the third axis ($a$). Therefore there are different
choices to be made in this method, namely, which is the initial set of
points $V$, the distance measure $d$ and the way in which $a$ is
defined. Different approaches have been followed to set these
quantities.

\begin{itemize}
\item \citet{Warren1992} use an iterative scheme keeping the ellipsoid
  volume constant. The set $V$, initially chosen to be a spherical
  shell, is iteratively deformed and reoriented using the eigenvalues
  and eigenvectors of the reduced inertia tensor. These authors take
  $d$ to be the Euclidean distance to a given particle.
  \citet{Bailin2005}, however, argue that this systematically
  overpredicts the roundness of the haloes. In the figures below we
  have modified \citeauthor{Warren1992}'s method by using $d_k^2 =
  x_k^2 + y_k^2/q^2 + z_k^2/s^2$ instead, and where $q=b/a$ and
  $s=c/a$ are updated in each iteration. We define also the shell's
  radii using this definition of distance, instead of the Euclidean
  one.

\item \citet{Allgood2006} also employ an iterative scheme but now
  keeping the largest axis length constant. Initially the set $V$ is
  selected to be given by all particles located inside a sphere (as
  opposed to a spherical shell of a given radius) which is reshaped
  iteratively using the eigenvalues. As before, the orientation is
  determined from the eigenvectors of $\vect{I}$, and the distance
  measure used is $d_k^2 = x_k^2 + y_k^2/q^2 + z_k^2/s^2$. In the
  figures below we have removed all bound substructures contained in a
  halo. This alleviates the noise and artificial tilting of the
  ellipsoids that is introduced by such substructures. We use
  \textsc{Subfind} \citep{Springel2001} to identify and remove
  particles associated to subhaloes within the region of interest.

\end{itemize}

In our implementations the iterations are stopped when convergence in
the axis ratios is reached, which we take to be when the relative
change is smaller than $10^{-6}$, or when $V$ is composed by less than
3000 particles. If the latter condition is not fulfilled the shape of
such a contour is not considered in our analysis.

\begin{figure}
  \centering \includegraphics[width =
  0.45\textwidth]{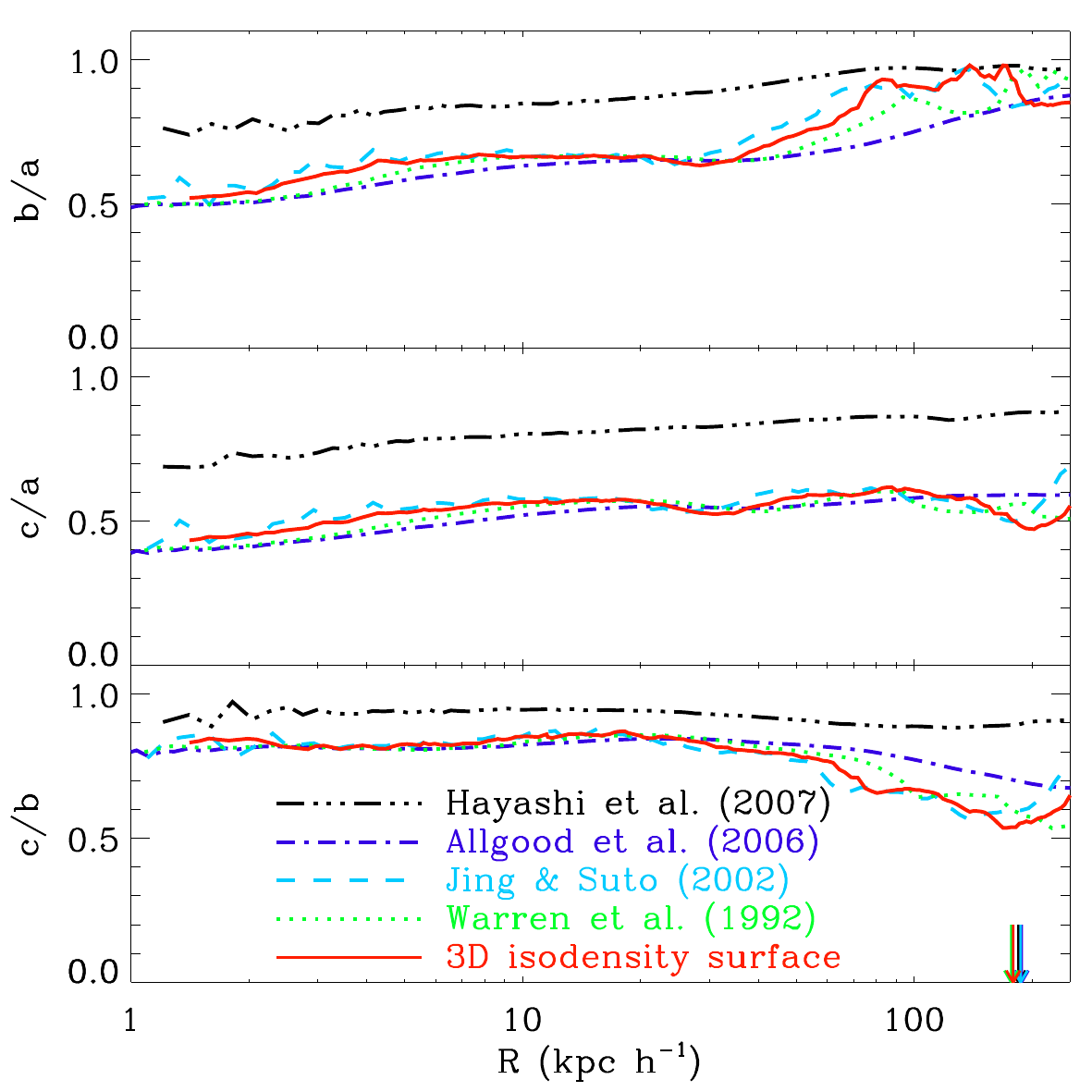}
  \caption{\label{axis-ratios-vs-R-other} Axis ratios ($a\geq b \geq
    c$) for \Aq{A-4} as a function of the ellipsoidal distance
    $R=(abc)^{1/3}$. The solid red line corresponds to our new
    algorithm, while dot-dashed black, dashed cyan, dot-dashed blue
    and dotted green correspond respectively, to: iso-potential
    contours \citep{Hayashi2007}, particle-based iso-density inertia
    tensor \citep{Jing2002}, normalized inertia tensor diagonalized
    upon hollow \citep{Warren1992} and on solid \citep{Allgood2006}
    ellipsoids. Vertical arrows indicate the size of the virial
    ellipsoid, $R_{\rm vir}$, for each method. All methods agree well,
    especially in the inner regions ($R\lesssim 100 h^{-1}$ kpc).}
\end{figure} 

\subsection{Density}

An alternative approach to determine halo shapes is to consider the
underlying density field, which carries more information about the
internal mass distribution of the haloes than the inertia tensor.  We
explore here two different implementations.

\begin{itemize}
\item \citet{Jing2002} determine the shape by fitting ellipsoids to
  sets of particles having (nearly) the same nearest neighbours-based
  density. Noise due to substructures is effectively removed by an
  implementation of the Friends of Friends (\textsc{fof}) algorithm to
  these sets \citep{Davis1985}, and where the linking length is
  selected to vary from set to set of iso-density particles according
  to the empirical law $l = 3 (\rho/m)^{-1/3}$.

\item We present a new method based on fitting an ellipsoid to the
  particle density. We first create a continuous density field out of
  the particles positions. We do so by using a Cloud-In-Cell algorithm
  that allows the reconstruction of the density field on a regular
  grid \citep{Hockney1988}. For better resolution and to keep the
  computational cost relatively low, the region covered by the grid is
  iteratively increased. The second step involves the identification
  of iso-density contours. We select the cells with nearly the same
  density and a version of the \textsc{fof} algorithm is used to get
  rid of cells associated to substructures artificially linked to the
  main contour. Finally, we minimize the function:
  
  \begin{equation}    
    \mathcal{S}(\vect{M}) = \frac{1}{n}\sum_{k=1}^n \left( 1 -
    \sqrt{\vect{x}_k^T \vect{M}\vect{x}_k}\right)^2,
  \end{equation}
  
  \noindent with $\vect{M}$ the matrix representation of an ellipsoid,
  in order to determine the axis lengths (eigenvalues of $\vect{M}$)
  and directions (eigenvectors). Notice that the minimization is
  carried out in a $6$ dimensional space ($\vect{M}$ has just $6$
  independent elements), therefore an educated initial guess for the
  iteration may reduce the numerical effort. We provide this guess by
  diagonalizing the inertia tensor of the cells with similar values of
  the density.

\end{itemize}

Fig.~\ref{density-contours} shows 2D slices of the density map
computed for halo \Aq{A-4}, together with several best-fit ellipsoids
found by the method just outlined. An important feature of our method
is that all the information about the 3D isodensity {\it surface} is
taken into account; this is particularly useful towards the outskirts
of the dark matter haloes where, as can be appreciated from
Fig.~\ref{density-contours}, density contours become less
symmetric. For comparison, we also show the results of applying the
algorithm by \citet{Allgood2006} (after subhaloes subtraction) in
dashed lines. Recall that in our new method ellipsoids are effectively
independent of each other and therefore this method is more sensitive
to local variations of the halo shapes. On the other hand,
\citet{Allgood2006} use the whole set of particles within a given
radius, which implies that the shape of a contour at a given distance
is correlated with the shape at smaller radii, as a careful inspection
of this figure shows.

\subsection{Potential}

A viable alternative to the density-based methods to measure the shape
of a dark matter halo is to use the gravitational potential field
defined by the particles. As first noticed by \citet{Springel2004} and
later confirmed by \citet{Hayashi2007}, substructures cause
significant fluctuations in the local density distribution, but their
contribution to the gravitational potential is considerable less
harmful \citep{Springel2004, Hayashi2007}. Iso-potential contours are
therefore smoother and more regular than those defined by the density
field. Taking advantage of this feature, \citet{Hayashi2007} have
implemented a method to characterize the structural shape of dark
matter haloes by fitting ellipses to the iso-potential contours
computed on three orthogonal 2D planes. It is important to note here
that the potential is intrinsically rounder than the density
distribution, for example for a cored logarithmic potential $1 -
(c/a)_\rho \approx 3[1 - (c/a)_\Phi]$ outside the core
\citep{Binney2008}.

\subsection{Results}

In Fig.~\ref{axis-ratios-vs-R-other}, we compare the shape as a
function of the ellipsoidal radius obtained for the halo \Aq{A-4}
using the methods described above. This figure shows very good
agreement in the halo shapes measured by the different methods,
especially in the inner regions ($R \lesssim 100 h^{-1}$ kpc). There
is however an indication that at large radii, density-based methods
(red solid and light-blue dashed curves) tend to give slightly more
oblate shapes (higher $b/a$) than those based on implementations of
the inertia tensor, but the effect is only marginal.  As expected, the
method based on the gravitational potential \citep{Hayashi2007}
produces higher axis ratios (blue dotted line).

\section{The size of filaments at different times}
\label{sec:filament_size}

\begin{figure}
  \centering \includegraphics[width =
  0.45\textwidth]{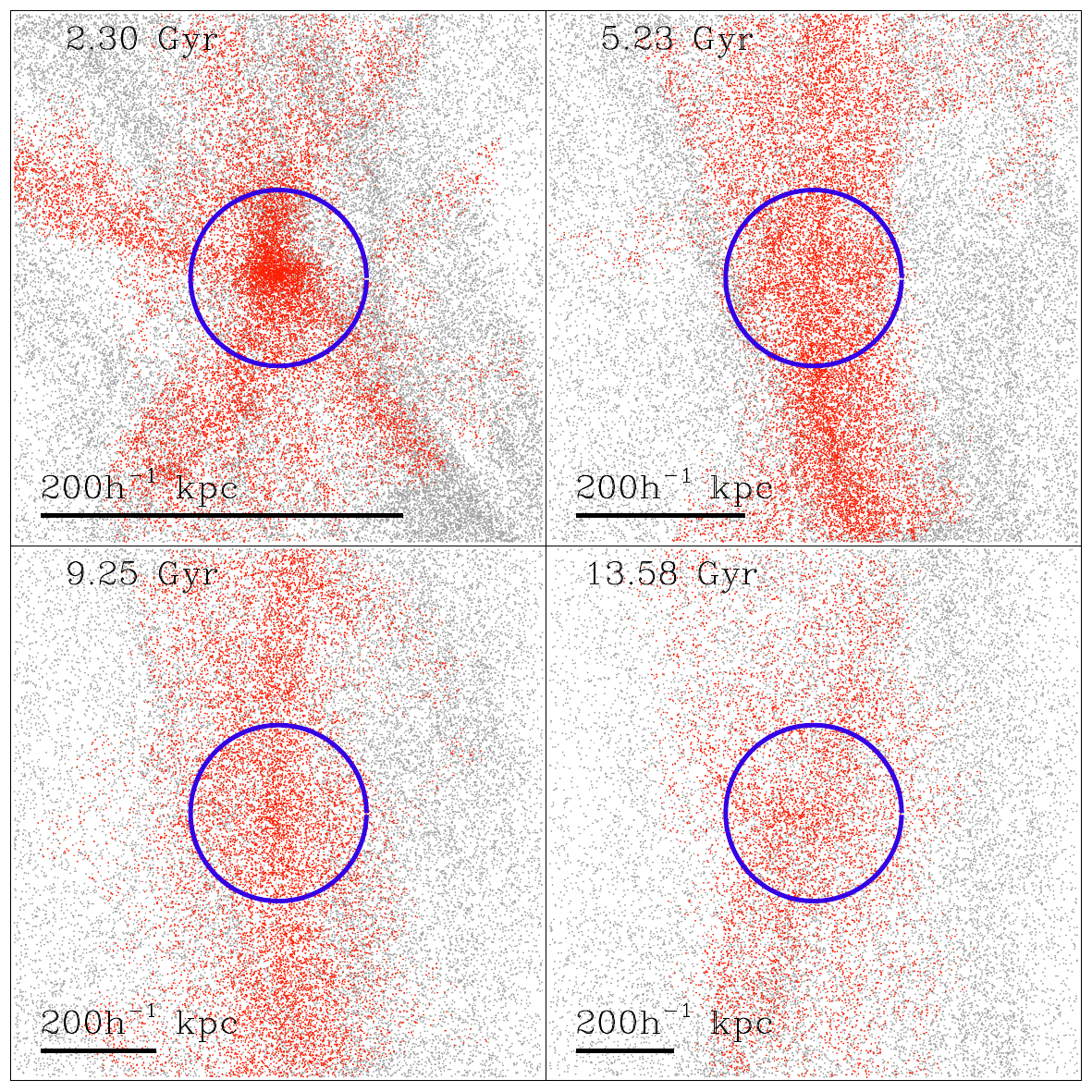}
  \caption{\label{fig:caustics1} Distribution of particles around
\Aq{A} halo at four different times. Particles with a single caustic
crossing are plotted in red and are seen to trace reasonably well the
filamentary structure surrounding the halo.  Each box has been rotated
according to the inertia tensor defined by this subset of particles.}
\end{figure} 

\begin{figure*}
  \begin{center} 
  \includegraphics[width = 0.30\textwidth]{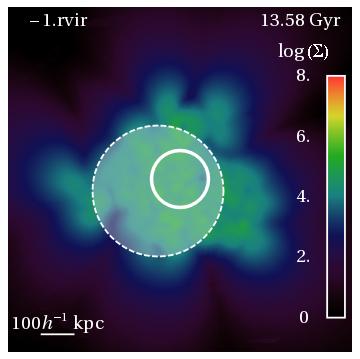}
  \includegraphics[width = 0.30\textwidth]{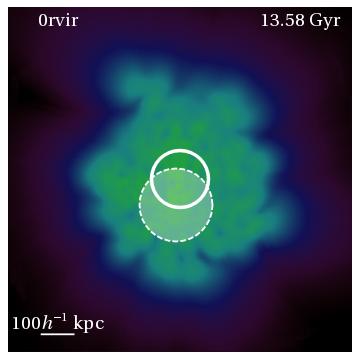}
  \includegraphics[width = 0.30\textwidth]{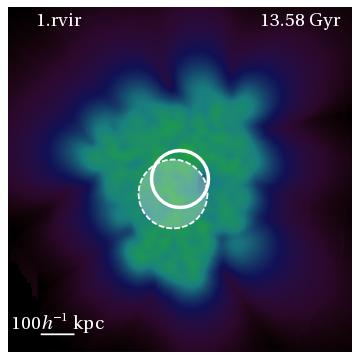} \\
  \includegraphics[width = 0.30\textwidth]{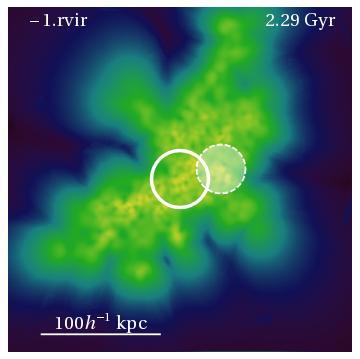}
  \includegraphics[width = 0.30\textwidth]{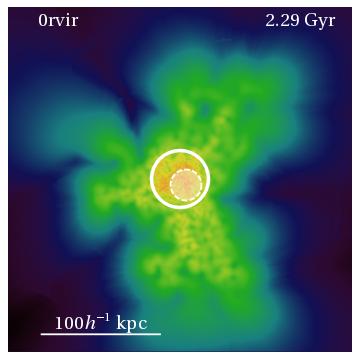}
  \includegraphics[width = 0.30\textwidth]{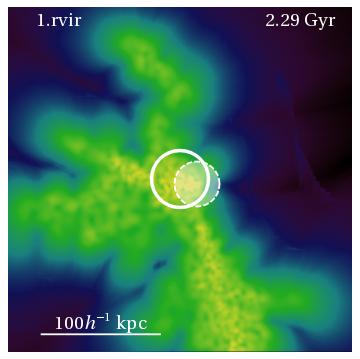}
  \end{center}
  \caption{\label{fig:caustics2} Projected mass density (in units of
    $h$M$_\odot \rm kpc^{-2}$) for three different perpendicular
    planes located along the direction of the filament at positions
    $z_p=[-1,0,1] r_{\rm vir}$, and at two different times: $t=13.6$
    Gyr (top) and $t=2.3$ Gyr (bottom).  The solid white curve
    indicates the half the virial radius of the halo at each time and
    the highlighted filled disc shows the size assigned to the
    filament and centred in the highest density point as described in
    the text. The geometry of the filaments is complicated, but their
    relative size with respect to that of the halo is clearly larger
    at present day ($t\sim13.6$ Gyr) than it was at $t\sim2.3$ Gyr.}
\end{figure*} 

Dark matter haloes that reside in a filament will acquire their mass
preferentially along the filament's longest axis. As discussed in
Sec.~\ref{sec:lss}, such infall of material gives rise to a power
spectrum characterized by a significant $l=2$ component. However, when
the surrounding filament is sufficiently wide, i.e. of comparable or
larger cross-section than the virial radius of the halo, the infalling
particles will appear to be more isotropically distributed on the sky,
shifting the power to the $l=0$ term of the spectrum. We have argued
in Section \ref{sec:lss} of this paper that the latter case is
characteristic of the late stages of mass assembly in $\sim
10^{12}$M$_\odot$ objects. We analyze this statement in more detail in
this Appendix, and provide a suitable measurement of a filament's size
and compare it to that of the dark matter halo it hosts. For brevity
we focus our analysis on halo \Aq{A}, but our conclusions should hold
in the general case.

The measurement of the size of a filament is not completely
straightforward and may depend on the particular algorithm used for
its identification \citep{Stoica2005, Zhang2009}. In this work we
define structures according to the number of caustic crossings of its
constituent particles. Caustics arise during the gravitational
collapse of a dynamical system. In the case of cold dark matter, the
initial velocities of particles are negligible, and thus these are
distributed in 3D sheets in phase-space \citep{Bertschinger1985}.
Their collapse and subsequent virialization may be seen as the folding
of these sheets, and the location of these folds in configuration
space gives rise to caustics. Therefore the number of caustic
crossings is indicative of the degree of virialization of a dynamical
system.

For example particles with zero crossings remain under the
quasi-linear regime and have therefore not collapsed into any
virialized structure today.  As gravitational collapse proceeds, the
number of caustics that a particle experiences increases rapidly. As
shown in \citet[][Fig. 4]{Vogelsberger2010}, particles with 1 or 2
caustic crossings delineate the surrounding filamentary structure of a
halo, whereas those with a higher number of crossings belong to the
host halo itself.  Therefore, in this work we select particles with 1
caustic crossing to study the properties of the filamentary structure
surrounding \Aq{A} halo at different epochs \citep{Vogelsberger2009,
Vogelsberger2010}.

Fig.~\ref{fig:caustics1} shows the distribution of particles in a box
of size $6r_{\rm vir}$ at four different times, where those with a
single caustic crossing are highlighted in red. The circle indicates
the corresponding virial radius of the \Aq{A} halo, and the horizontal
bar gives the reference for conversion to the physical scale. At each
time-step we have rotated the reference frame to the principal axis of
the inertia tensor defined by the particles with 1 caustic crossing
\footnote{Because we are interested in the set of filaments connected
  to the central object, we run a \textsc{fof} algorithm over the
  particles with 1 caustic crossing, using a linking length of 0.7 the
  mean interparticle separation and retain for the analysis only those
  set of particles belonging to the most massive \textsc{fof}
  group}. This figure shows these particles successfully trace the
filamentary structure around this halo. Interestingly, there is a hint
of evolution in the relative size of the filament with respect to that
of the halo: as we move back in time, the red particles change from
fully encompassing the halo ($t\geq 9$ Gyr) to having a similar
cross-section ($t \sim 5.25$ Gyr) to becoming even narrower at earlier
times ($t \sim 2.3$ Gyr). At this point the geometry of the system is
much more complex due to multiple filaments feeding material to the
central object.

In order to quantify the evolution in the cross section of a filament
we proceed as follows. We rotate the reference system such that the
$z$-direction is that given by the inertia tensor of particles with 1
caustic crossing. We identify planes perpendicular to this direction
and label the positions of such planes as $z_p$, where
$z_p=[-2,-1.5,-1,...,+1.5,+2]$ $r_{\rm vir}$. We project in each plane
all selected particles that satisfy $|z_i-z_p| \leq 0.5r_{\rm vir}$.
We compute the (projected) density of neighbours of each particle in
the plane using a 2D SPH kernel and identify the particle with the
highest density.  This particle's location sets the centroid of the
filament and its density at the core. We then define the radius of the
filament $r_{\rm fil}$ as the distance from this centroid where the
density has dropped to $60\%$ of its central value. Although this
definition is arbitrary, this allows a measurement of the relative
size at different times.

Fig.~\ref{fig:caustics2} illustrates our procedure for two different
times: $t=13.6$ (top) and $t=2.3$ (bottom) Gyr; applied to 3 different
planes along the $z$-direction: $z_p=-r_{\rm vir},0,+r_{\rm
vir}$. Each panel shows the projected density of particles belonging
to the filamentary structure (one single caustic crossing), where the
virial radius of the central halo is indicated with a white (empty)
circle, and the highlighted full disc indicates the size of the
filament centred on the highest density point as described above.
Fig.~\ref{fig:caustics2} suggests that filaments are not really
straight in space (as indicated by the different positions of the
centroids for panels located at different $z_p$) and also their
geometry is complex since panels located symmetrically with respect to
the centre of the halo yield significantly different sizes (e.g. top
row, $-r_{\rm vir}$, $+r_{\rm vir}$ panels). However, there is still a
clear trend indicating that the filament's relative size was smaller
at early times than at the present day.

\begin{figure}
  \centering
  \includegraphics[width = 0.45\textwidth]{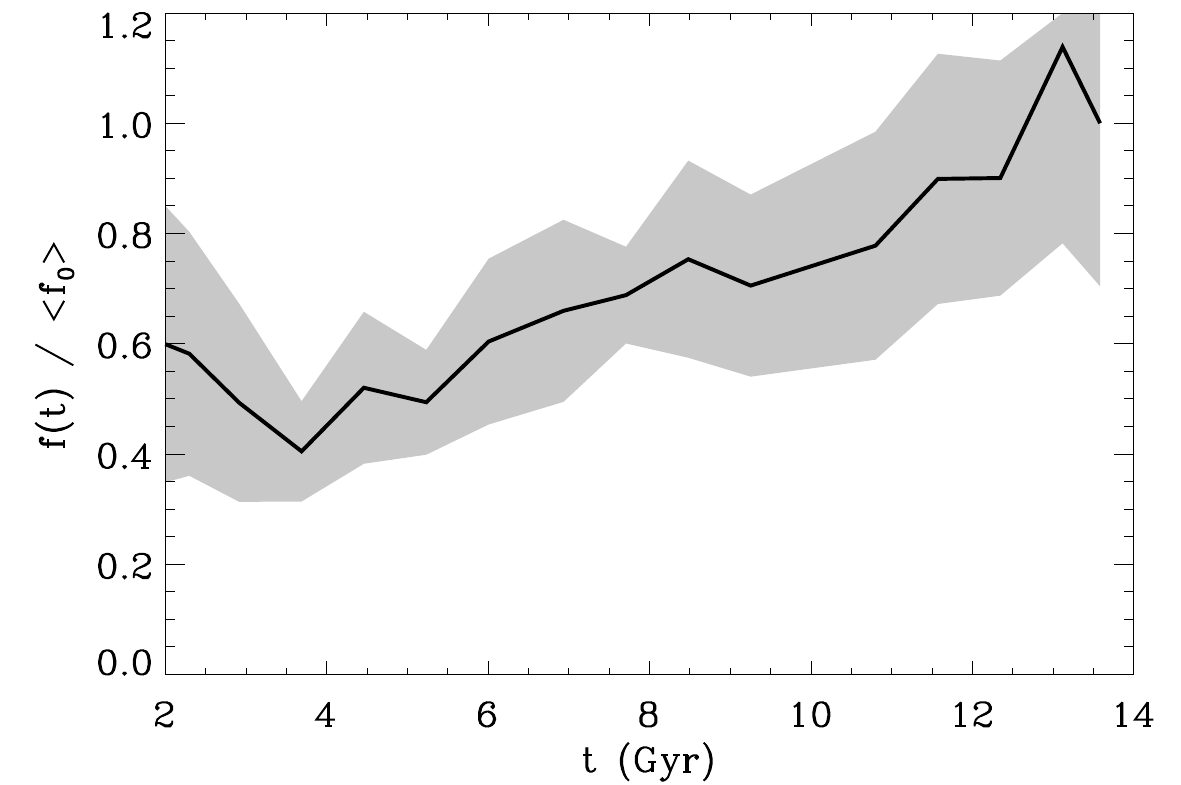}
  \caption{\label{fig:caustics3} Filament size relative to the virial
    radius $f=r_{\rm fil}/(0.5r_{\rm vir})$ as a function of time. The
    solid black line is the average size of the filament measured on 9
    planes perpendicular to the direction of the filament. The shaded
    region is $1\sigma$ scatter (see text for details) . The vertical
    axis has been normalized to its final value at redshift zero
    $f_0=f(z=0)$ for an easier comparison.}
\end{figure}

This trend is more clearly seen in Fig.~\ref{fig:caustics3}, where we
show the average size of the filament (normalized to a fraction of the
instantaneous virial radius) measured as the average over 9 equally
distant planes located from $[-2 r_{\rm vir}, +2 r_{\rm vir}]$ as a
function of time.  The shaded region shows the scatter between the
sizes derived for each of the individual planes at a fixed time.  Notice
that the relative filament-to-halo size at present day is almost
twice as large than its value at $t \sim 2$ Gyr. This provides further
support to our observation that the infall of particles at later times
results from an increased size of the filament with respect to that of
the halo.

\end{document}